\documentclass[12pt]{article}
\usepackage{ascmac}
\usepackage{ulem}
\usepackage{latexsym}
\usepackage[dvipdfmx]{graphicx}
\usepackage{amsmath}
\usepackage[top=3.5cm,bottom=3.5cm,left=2cm,right=2cm]{geometry}
\usepackage{pifont}          
\usepackage{amsmath,amssymb}
\usepackage{color}
\usepackage{geometry}
\usepackage{caption}
\usepackage{ulem}
\usepackage{cancel}
\newcommand{\nn}{\nonumber\\}
\newcommand{\h}{\hspace}
\newcommand{\be}{\begin{equation}}
\newcommand{\e}{\end{equation}}
\newcommand{\aln}[1]{\begin{align}#1\end{align}}
\begin{document}
\title{
\vbox{
\baselineskip 14pt
\hfill \hbox{\normalsize KEK-TH-2071}} 
\vskip 1cm
\bf \Large  Density Renormalization Group for Classical Liquids  \\
\vskip 0.5cm
}
\author{
Satoshi Iso$^{a,b}$\thanks{E-mail: \tt satoshi.iso(at)kek.jp}\h{1mm} 
and Kiyoharu~Kawana$^{a}$\thanks{E-mail: \tt kawana(at)post.kek.jp}
\bigskip\\
\it 
\normalsize
 $^a$ Theory Center, High Energy Accelerator Research Organization (KEK), \\
 \normalsize
\it  $^b$Graduate University for Advanced Studies (SOKENDAI),\\
\\
 \normalsize
\it Tsukuba, Ibaraki 305-0801, Japan \\
\smallskip
}
\date{\today}

\maketitle

\abstract{\normalsize
%
We study response of liquid to a scale transformation, 
which generates a change of the liquid density, and obtain a set of differential equations for correlation functions. 
The set of equations, which we call density renormalization group equations (DRGEs), is similar to the BBGKY hierarchy as it relates different multiple-point correlation functions.   In particular, we derive
DRGEs for one-particle irreducible vertex functions of liquid by performing Legendre transformations, which
enables us to calculate properties of liquid at higher density 
in terms of correlation functions at lower density.  
}
\newpage

\section{Introduction}
Thermodynamical properties of gas and liquid are described by the equation of state, and the microscopic
derivation is one of the most important issues in the liquid theory. 
Liquid is microscopically described by a collection of interacting particles and  the thermodynamical quantities such as pressure,  internal energy, and isothermal compressibility
can be calculated from  density correlation functions \cite{review1,review2}. 

Statistical mechanics of liquids has a long history. 
The simplest microscopic study of liquid is the virial expansion method, which gives a systematic expansion around the ideal gas. But its convergence is very slow \footnote{
For example, in the case of the hard-sphere potential 
whose height is $+\infty$ for $r<b$ and zero otherwise, the virial expansion becomes \cite{Hard Sphere}
\aln{\frac{p}{nT}=1+4\eta+10\eta^2+18.36\eta^3+28.22\eta^4+\cdots,
}
where $\eta$ is the packing fraction defined by $\eta=\frac{4\pi b^3 N}{3}/V$.
%
For example, setting $\eta=0.3$, the contribution of the fifth term  is as large as $\sim 30/80\sim 0.37$. 
} and the method is limited to low density region. 
In order to describe high density region of liquid, we need to go beyond the virial expansion and take into account  effects of strong multi-point density correlations. 
For this purpose, various integral equations and their approximations have been studied. 
The Orstein-Zernike equation, which is an integral equation for two-point correlation functions (see Eq. (\ref{eq:OZ})),
is often used and various approximations such as the Percus-Yevick approximation \cite{PY,PY-Hard1,PY-Hard2,PY-LJ1,PY-LJ2} or the Hyper-Netted chain approximation \cite{HNC1,HNC2,HNC3,HNC4,HNC5} are proposed. 
Another type of integral equations is the BBGKY hierarchy \cite{BBGKY1,BBGKY2,BBGKY3,BBGKY4,BBGKY5}, which 
is a set of integral equations relating different multi-point correlation functions. 
Thus, in order to solve them, we need to cut the chain of equations at some orders. 
In the Kirkwood superposition approximation \cite{BBGKY3,BBGKY4,Kirkwood1,Kirkwood2}, 3-point correlation function is assumed to be expressed in terms of a product of 2-point correlation functions so that the hierarchical equations are closed. 
Various types of equations and  approximations for closures have been proposed
and the results are in good agreement with Monte Carlo simulations of liquid \cite{Ashcroft, Hus}. 
See also \cite{review1,review2} and references therein for more details. 
In this paper, based on the method \cite{Hubbard}, 
we propose another type of differential-integral equations which describes
response of physical quantities to a change of the liquid density, and thus
enables us to calculate the liquid's property at higher density from the quantities at lower one.

The present work is motivated  by the similarity with
the renormalization group equation (RGE) in quantum field theories (QFT).
RGEs tell us the energy dependence of various  physical quantities in QFTs, such as an effective coupling;
just as an effective coupling 
varies as a function of the energy scale (i.e. renormalization scale) in QFT,
pressure of the liquid changes as a function of liquid density. 
In both theories,  physical quantities are calculated from correlation functions. 
Thus we can infer the following analogies between classical liquid theories and quantum field theories:
\begin{itemize}
\item Helmholtz free energy $\longleftrightarrow$ Effective potential  $\Gamma$
\item Equation of state  $\longleftrightarrow$   RG improved equation of motion for $\Gamma$
\item  Pressure, compressibility, etc.  $\longleftrightarrow$ Running couplings
\item  Density $\longleftrightarrow$ Renormalization scale $\mu$
\end{itemize}
Such analogies between statistical mechanics and quantum field theories have been occasionally pointed out, 
 but our present study is strongly stimulated by Nambu's seminar paper \cite{Nambu} in which an analogy between the renormalization group (RG) equation in 
gauge theories and thermodynamic equation of state was discussed. 
See also the seminal papers \cite{Parola1,Parola2,Caillol} and the references therein where the ordinary concept
of the RG is applied in the classical liquid/vapor system.\footnote{
The studies of these paper \cite{Parola1,Parola2,Caillol}  are similar to ours in that 
 both approaches are based on the hierarchy of coupled integro-differential equations for the density correlation
 functions. But there is a big difference.
 In their studies, the hierarchical equations are obtained by gradually including the 
 attractive part of the interaction potential as the IR cutoff is changed, and  
 the approach of \cite{Parola1,Parola2,Caillol} is closer to the original idea of the  
 Wilsonian renormalization group scheme. 
 Therefore it is powerful near the critical point where the system becomes 
 scale invariant and the density fluctuation is anomalously enhanced.
On the other hand, our hierarchical equation is obtained not by changing the IR cutoff but by 
gradually including the effects of liquid density. Thus  the present formalism is expected to be
useful in the normal state at higher density rather than near the critical point. 
}

The purpose of the present paper is to make the analogy more concrete and to 
propose a set of (density) differential equations for correlation functions by studying
response of the liquid to a scale transformation.
Since a scale transformation  generates a change of liquid density,
the resultant equations describe response of  correlation
functions to a small change of density; thus we call them density renormalizaton group 
equations (DRGEs). 
The physical meaning of DRGEs are the following.
In (classical) gas and liquid, when its density is low, the system is well described by
the ideal gas, and the density correlation functions are exactly calculated (see \ref{app:ideal}).
As the density increases,  correlation functions start to behave nontrivially due to 
two different reasons. 
One is, of course, a direct consequence of intermolecular (2-body) potential between particles.
This causes a nontrivial behavior for the 2-point 
density correlation function. But there is another effect. 
Nontrivial (multi-) correlation (more than 2-point functions) will appear
due to finite density effects. Namely, e.g. 
density fluctuations at 3 different points get correlated  mediated by
finite density effect at a single point in the middle. 
This effect becomes stronger in higher density liquid and also when 2-point correlation becomes stronger.  
Therefore, if we can resum (or accumulate) these effects from low to high density,
we will be able to describe the dynamics of high density liquid by solving
DRGEs. 

The paper is organized as follows. 
In Section \ref{sec:review}, we first briefly review the statistical mechanics of  classically interacting particles
and then study how a partition function of such a system responds to a scale transformation.  
In this way, we derive  a set of partial differential equations for correlation functions.
In Section \ref{sec:hierarchical}, by using a field theoretical method by Hubbard and Schofield, 
we calculate perturbative corrections to correlation functions 
by a small change of density and 
obtain  explicit forms of the density renormalization group equations (DRGEs). 
We also briefly mention how we can solve the DRGEs to obtain thermodynamical behaviors of liquid in \ref{app-solve}.
Section \ref{sec:Helmholtz} is the main part of the paper. 
We perform Legendre transformations to obtain the Helmholtz free energy
and derive another type of DRGEs. 
Like the analysis in Section \ref{sec:hierarchical}, the calculations are based on the diagrammatic methods.
We will see that the Helmholtz free energy generates one-particle irreducible (1PI) diagrams; thus the correlation functions generated by the Helmholtz free energy correspond to the 1PI vertices in QFT. 
 For a consistency check, we confirm that our DRGEs for 1PI vertices correctly reproduce the ordinary results of Mayer's cluster expansion up to the third virial coefficients in \ref{app:virial}.   
Section \ref{sec:summary} is devoted to summary and discussion. 

\section{ Response to Scale Transformations}\label{sec:review}
In this section, we investigate how  classical liquid (or gas), i.e. a set of classically interacting particles, responds to scale transformations.  
A scale transformation generates a change of the liquid density and accordingly  
we can obtain partial differential equations  describing how the system changes according to the change of the density.   

In section \ref{subsec:review}, in order to fix our notations we briefly review  various properties of classically interacting particles, and then in section \ref{subsec:scale} we obtain the DRGEs for density correlation functions. 

\subsection{Brief review of classically interacting particles}
\label{subsec:review}
We consider statistical mechanics of $d$-dimensional
 classically interacting particles whose Hamiltonian is given by 
\aln{
H_N^{}=\sum_{i=1}^N\frac{p_i^2}{2m}+\sum_{i<j}^Nv(x_i^{},x_j^{}),
}
where $v(x,y)$ represents two-body interactions. 
In this paper we assume that  $v(x,y)$ is a function of the relative distance 
$v(x,y)=v(|x-y|)$, which reflects the translational symmetry of the system and the absence of
polarizations (i.e. simple liquids). 
Furthermore we neglect many-body interactions for simplicity.

The grand-canonical partition function is given by
\aln{
\Xi_v^{}[T,\mu,V;U]
 &=\sum_{N=0}^\infty \frac{e^{\beta \mu N}}{N!}\int_V d^dx_1^{}\int d^dp_1^{}\cdots
\int_V d^dx_N^{}\int d^dp_N^{}\exp\left(
-\beta H_N^{}+\beta \sum_{i=1}^N U(x_i^{})
\right)
\nn
&=\sum_{N=0}^\infty \frac{z(\mu)^N}{N!}\int_V d^dx_1^{}\cdots
\int_V d^dx_N^{}\exp\left(
-\beta\sum_{i<j}v(x_i^{}-x_j^{})+\beta\sum_{i=1}^N U(x_i^{}) 
\right)
,
}
where $\beta=1/T$ is the inverse temperature and $z(\mu)$ is defined by
\be z(\mu)=e^{\beta\mu}(2\pi mT)^{d/2} .
\e
We have introduced the external source $U(x)$ for later convenience.  
By using the density operator 
\aln{\rho(x)\equiv \sum_{i=1}^N\delta^{(d)}(x-x_i^{}),
}
the grand-canonical partition function can be rewritten as
\aln{
\Xi_v^{}[T,\mu,V;U]&= \sum_{N=0}^\infty \frac{z(\mu+v(0)/2)^N}{N!}\int_V d^dx_1^{}\cdots
\int_V d^dx_N^{}\exp\left(
-\frac{\beta}{2}\langle \rho|v|\rho\rangle+\langle \beta U | \rho\rangle
\right)
,\label{eq: partition 2}
}
where
\aln
{
& \langle \rho|v|\rho\rangle\equiv \int d^dx\int d^dy\rho(x)v(x-y)\rho(y),  \\
& \langle \beta U | \rho\rangle=\beta\int d^dx U(x) \rho(x).
}
The functional derivatives of the grand  potential, 
\be -\beta W_v^{}[T,\mu,V;U] = \log \Xi_v^{}[T,\mu,V;U] ,
\e 
with respect to $\beta U(x)$ produce  correlation functions of the density fluctuation $\delta \rho(x)=\rho(x)-\langle \rho(x)\rangle$:
\aln{
&F_1^{}(x)\equiv \frac{\delta (-\beta W_v^{}[T,\mu,V;U])}{\delta (\beta U(x))}\bigg|_{U=0}^{}=\langle\rho(x)\rangle\equiv n,
\nn 
& F_l^{}(x_1^{},\cdots,x_l^{})\equiv \frac{\delta^l (-\beta W_v^{}[T,\mu,V;U])}{\delta(\beta U(x_1^{}))\cdots \delta(\beta U(x_l^{}))}\bigg|_{U=0}^{}=\langle\delta\rho(x_1^{}) \cdots \delta\rho(x_l^{})\rangle \ (\text{for }l\geq 2).
}
Here we assumed that the translational symmetry is not spontaneously broken at $U(x)=0$
and the particle density $\langle\rho(x)\rangle$  does not depend on its position $x$, which 
is written as $n$. 
Note that these correlation functions $F_l$ correspond to the connected 
Green functions  in the Hubbard's field theoretic formulation of classical liquids \cite{Hubbard}. 

The correlation functions are related to the thermodynamical quantities such as the isothermal
compressibility. In order to see this,
we write the density in presence of the external source $U=\Delta U(x)$ as
$n(x)$ and expand $\Delta n(x)=n(x)-n$ with respect $\Delta U(x)$; 
\aln{\Delta  n(x) 
&=\int d^dy\frac{\delta n(x)}{\delta U(y)}\bigg|_{\Delta U=0}^{}\Delta U(y)+\frac{1}{2} \int d^dy\int d^dz\frac{\delta^2 n(x)}{\delta U(y)\delta U(z)}\bigg|_{\Delta U=0}^{}\Delta U(y)\Delta U(z)+\cdots 
\nn
&=\int d^dyF_2^{}(x,y)\beta \Delta U(y)+\frac{1}{2} \int d^dy\int d^dzF_3^{}(x,y,z)
(\beta\Delta U(y))(\beta \Delta U(z))+\cdots.
}
Setting $\Delta U(x) =\Delta \mu=const$, 
we have 
\aln{\Delta n=\beta\Delta \mu\int d^dyF_2^{}(x,y)+\frac{(\beta \Delta \mu)^2}{2} \int d^dy\int d^dzF_3^{}(x,y,z)+\cdots,
}
from which we obtain the following relations between the thermodynamical quantities and 
integrals of the correlation functions;
\aln{
\kappa_T^{}\equiv &\frac{1}{n}\frac{\partial n}{\partial (\beta\mu)}\bigg|_{T,V}^{}=\frac{1}{n}\int d^dyF_2^{}(x,y),
\label{eq: compressibility by F2}
\\
\lambda_l^{}\equiv &
\frac{1}{n}\frac{\partial^{l-1} n}{\partial (\beta\mu)^{l-1}}\bigg|_{T,V}^{}=\frac{1}{n}\int \cdots \int 
\prod_{i=2}^l d^dx_i  \ F_l^{}(x_1, \cdots ,x_l),\ (\text{for }l\geq 3)
\label{eq: higher derivative and Fl}
}
where $\kappa_T^{}$ is called the ``{\it isothermal compressibility}" because it can be rewritten as
\aln{\kappa_T^{}=T\frac{\partial n}{\partial p}\bigg|_{T,V}^{}
\label{kappaT}
}
by using the thermodynamical relations.\footnote{When $T$ and $V$ are fixed, the differential form of the grand potential becomes
\aln{dW=-pdV+SdT-Nd\mu=-Nd\mu.
}
Because of  $W=-pV$, the relation $Vdp=Nd\mu$ follows for fixed $T$ and $V$.
Thus we have
\aln{
\frac{\partial }{\partial p}\bigg|_{T,V}^{}=\frac{1}{n}\frac{\partial }{\partial \mu}\bigg|_{T,V}^{}.
}
which relates Eq. (\ref{eq: compressibility by F2}) and  Eq. (\ref{kappaT}).
}
It is an indicator of the response of the fluid density against a small change of the external pressure. 
The first equation (\ref{eq: compressibility by F2}) is well known as the ``{\it isothermal compressibility equation}". 
It is usually written as the following form, 
\aln{\kappa_T^{}=1+n\int d^dx\left(\frac{1}{n^2}n^{(2)}(x,0)-1\right)=1+n\int d^dx h_2(x,0),
\label{eq: compressibility by h2}
} 
where  the ``two-point distribution function" $n^{(2)}(x,y)$ and the ``total correlation function" $h_2^{}(x,y)$ are defined by
\aln{n^{(2)}(x,y)-n^2\equiv n^2 h_2^{}(x,y)\equiv F_2^{}(x,y)-n\delta^{(d)}(x-y).
\label{eq:total correlation function}
}
Note that the total correlation function $h_2(x,y)$ vanishes for the ideal gas (see \ref{app:ideal}), and therefore $\kappa_T=1$. 

We can similarly define the ``$l$-point distribution function"  $n^{(l)}$, and the
``$l$-point total correlation function" $h_{l}$ 
for higher $l$ by 
\aln{n^{(l)}(x_1^{},\cdots,x_l^{})&\equiv n^l \left(h_l^{}(x_1^{},\cdots,x_l^{})+1 \right)
\nn
&= \left\langle\sum_{i_1^{}}\sum_{i_2^{}\neq i_1^{}}\cdots \sum_{i_l^{}\neq i_{l-1}^{}\neq\cdots i_1^{}}\delta^{(d)}(x_1^{}-x_{i_1^{}}^{})\delta^{(d)}(x_2^{}-x_{i_2^{}}^{})\cdots \delta^{(d)}(x_l^{}-x_{i_l^{}}^{})\right\rangle .
}
For an ideal gas, $h_l=0$ and thus $\lambda_l^{}=1$. 

%

The relations in Eq. (\ref{eq: compressibility by F2}) and Eq. (\ref{eq: higher derivative and Fl}) tell us that various thermodynamical quantities are expressed as integrals of the correlation functions. 
In order to evaluate these integrals, various approximations have been exploited such as the Kirkwood superposition assumption \cite{BBGKY3,BBGKY4,Kirkwood1,Kirkwood2}, the Percus-Yevick approximation \cite{PY,PY-Hard1,PY-Hard2,PY-LJ1,PY-LJ2}, or the Hyper-Netted chain approximation \cite{HNC1,HNC2,HNC3,HNC4,HNC5}.  

These approaches based on integral equations 
can be applicable to high density regions beyond the ordinary virial expansion method. 
It  is, however, difficult to treat them analytically and most studies have relied on 
numerical computations with various approximations whose validity are not well understood. 
In the studies of classical liquids, we often ask the following questions:  
How does the pressure or the isothermal compressibility vary as a function of the liquid density? 
How does the phase transition  or the critical phenomena occur as we change the density?  
In order to answer these questions, various systematic formulations and approximations
have been proposed \cite{review1, review2}. 
In order to
understand the evolution of the thermodynamical quantities 
as an increase of the liquid density, 
we notice a similarity to the renormalization group (RG) in quantum field theories (QFTs).
In both cases,  we are interested in
 the response of a system against scale transformations. 
But there is a big difference. 
In quantum field theory at zero temperature and zero (e.g. baryon) density, scale transformations
induce a change of the energy scales; the renormalization scale $\mu$ is changed.
On the other hand, in classical liquids at nonzero temperature and nonzero density, the transformations 
induce a change of the magnitude of the liquid density. 
Therefore scale transformations in liquid theory lead to 
differential equations to describe response of thermodynamical quantities
against a change of the density $n$ (or the chemical potential $\mu$) instead of the renormalization 
scale.\footnote{
Amusingly both of the renormalization scale
and the chemical potential are denoted by the same symbol 
$\mu$.}

%

\subsection{Scale transformations 
}\label{subsec:scale}
In order to investigate response of the system against a scale transformation, 
we consider a transformation of the two-body  potential: $v(x)\rightarrow v(ax)$.  
Then the partition function changes as 
\aln{
\Xi_{v(ax)}^{}&[T,\mu,V;U(x)] \nn
&=\sum_{N=0}^\infty \frac{z(\mu)^N}{N!}
\int_V  d^dx_1^{}\cdots
\int_V d^dx_N^{}\exp\left(
-\beta\sum_{i<j}v(a(x_i^{}-x_j^{}))+\beta\sum_i U(x_i^{}) \right)
\nn
&=\sum_{N=0}^\infty \frac{z(\mu -dT\log a)^N}{N!}
\int_{a^dV} d^dx_1^{}\cdots
\int_{a^dV} d^dx_N^{}\exp\left(
-\beta\sum_{i<j}v(x_i^{}-x_j^{})+\beta\sum_i U(x_i^{}/a)\right)
\nn
&=\Xi_{v}^{}[T,\mu-dT\log a,a^dV;U(x/a)] .
}
Therefore we see that the change of the potential under the scale transformation,
$v(x) \rightarrow v(ax)$, 
is equivalent to the changes of the chemical potential $\mu$, the volume of the system $V$, 
and the external source $U(x)$. 
For an infinitesimal scale transformation $a=1+\epsilon$, we have 
\aln{\Xi_{v+\epsilon\delta^{} v}^{}[T,\mu,V;U(x)]=\Xi_v^{}[T,\mu-dT\epsilon,(1+d\epsilon)V;U(x(1-\epsilon))]}
or equivalently
\aln{
-\beta W_{v+\epsilon\delta^{} v}^{}[T,\mu,V;U(x)]
=-\beta W_v^{}[T,\mu-dT\epsilon,(1+d\epsilon)V;U(x(1-\epsilon))
] ,
\label{eq:relation1}
}
where  $\delta v(x)= x^\mu\partial_\mu^{}v(x)$. 
Thus, by differentiating it with respect to $\beta U(x)$, we obtain the following relation
of the correlation functions, \footnote{ The derivation is the following;  
By taking functional derivatives $l$ times, we obtain
\aln{
\text{LHS}&=F_l^{}(x_1^{},\cdots,x_l^{};\mu,V)+\epsilon \Delta  F_l^{}(x_1^{},\cdots,x_l^{};\mu,V)
\\
\text{RHS}&=\frac{\delta^l\log\Xi_v^{}[T,\mu-dT\epsilon,(1+d\epsilon)V;v(x),U(x(1-\epsilon))]}{\delta (\beta U(x_1^{}))\cdots \delta (\beta U(x_l^{}))}
\nn
&=\left(\prod_{i=1}^l\int d^dy_i^{}
\frac{\delta (U(y_i^{})-\epsilon y_i^\mu\partial_\mu^{}U(y_i^{}))}{\delta U(x_i^{})}\right)F_l^{}(y_1^{},\cdots,y_l^{};\mu-dT\epsilon,V+d\epsilon V)
\nn
&=F_l^{}(x_1^{},\cdots,x_l^{};\mu-dT\epsilon,V+d\epsilon V)
+\epsilon\sum_{i=1}^l \partial_{i\mu}^{}(x_i^\mu
F_l^{}(x_1^{},\cdots,x_l^{};\mu,V))
 \nn
 &=F_l^{}(x_1^{},\cdots,x_l^{};\mu-dT\epsilon,V+d\epsilon V)
+dl\epsilon F_l^{}(x_1^{},\cdots,x_l^{};\mu,V)+\epsilon \sum_{i=1}^l x_i^\mu
 \partial_{i\mu}^{}F_l^{}(x_1^{},\cdots,x_l^{};\mu,V) .
}
}
\aln{
& F_l^{}(x_1^{},\cdots,x_l^{})+\epsilon \Delta  F_l^{}(x_1^{},\cdots,x_l^{}) \nonumber \\
& =F_l^{}(x_1^{},\cdots,x_l^{})|_{\mu-dT\epsilon,V+d\epsilon V}^{}
+dl\epsilon F_l^{}(x_1^{},\cdots,x_l^{})+\epsilon \sum_{i=1}^l x_i^\mu
 \partial_{i\mu}^{}F_l^{}(x_1^{},\cdots,x_l^{}),
}
where $\Delta F_l^{}(x_1^{},\cdots,x_l^{})$ represents perturbative corrections due to $\delta v(x)$,
which are explicitly evaluated in the next section.
Finally, by taking the $\epsilon\rightarrow 0$ limit, we obtain the following set of
partial differential equations for the correlation functions:
\aln{d\left(-\frac{\partial}{\partial(\beta \mu)}\bigg|_{T,V}^{}+\frac{\partial}{\partial\ln V}\bigg|_{\mu,T}^{}+l+\frac{1}{d}\sum_{i=1}^lx_i^\mu\partial_{i\mu}^{}
\right)F_l^{}(x_1^{},\cdots,x_l^{})
=\lim_{\epsilon\rightarrow 0}\Delta F_l^{}(x_1^{},\cdots,x_l^{}).
\label{eq: PDE1}
}
The lhs  contains a derivative with respect to the chemical potential $\mu$. Thus
they describe how the classical liquids respond to a change of the chemical potential. 

By performing the Fourier transform
\aln{\tilde{F}_l^{}(k_1^{},\cdots,k_l^{})=\int d^dx_1^{}\cdots \int d^dx_l^{}e^{-i\sum_{i=1}^lk_i^{}\cdot x_i^{}}F_l^{}(x_1^{},\cdots,x_l^{}),
}
we have
\aln{d\left(-\frac{\partial}{\partial(\beta \mu)}\bigg|_{T,V}^{}+\frac{\partial}{\partial\ln V}\bigg|_{\mu,T}^{}-\frac{1}{d}\sum_{i=1}^lk_i^\mu\frac{\partial}{\partial k_i^\mu}\right)\tilde{F}_l^{}(k_1^{},\cdots,k_l^{})
=\lim_{\epsilon\rightarrow 0}\Delta \tilde{F}_l^{}(k_1^{},\cdots,k_l^{}).
\label{eq: PDE2}
}
Furthermore, because the chemical potential and the density are related each other through Eq. (\ref{eq: compressibility by F2}), Eq. (\ref{eq: PDE2}) can be also written by  
\aln{d\left(-\kappa_T^{} \frac{\partial}{\partial\ln n}\bigg|_{V,T}^{}+\frac{\partial}{\partial\ln V}\bigg|_{\mu,T}^{}
-\frac{1}{d}\sum_{i=1}^lk_i^\mu\frac{\partial}{\partial k_i^\mu}\right)\tilde{F}_l^{}(k_1^{},\cdots,k_l^{})=\lim_{\epsilon\rightarrow 0}\Delta \tilde{F}_l^{}(k_1^{},\cdots,k_l^{}),
\label{eq: PDE3}
}
which describes how the system (in particular its correlation functions) changes as we change the density. 
In the next section, we explicitly evaluate the corrections
 $\Delta F_l^{}$ to the correlation functions due to  change of the potential $\delta v(x) = x^\mu \partial_\mu v(x)$.
We adopt the field theoretical  approach to the classical interacting particles proposed by J. Hubbard and P. Schofield \cite{Hubbard}, and 
we will see that the corrections $\Delta F_l^{}$ 
are written in terms of  higher-body correlation functions such as $F_{l+1}^{}$.
Thus the above equations generate a hierarchical structure similar to the BBGKY hierarchy.
Note that the volume derivative at fixed $(\mu, T)$ vanishes in the large $V$ limit.
Similar, but different, hierarchical equations are given in \cite{Gray}. 
%

\section{Density Renormalization Group Equations
}\label{sec:hierarchical}

\subsection{
Field theory of classically interacting particles}\label{subsec:Hubbard}
One of the successful perturbative approaches in the classical liquid/vapor theory is the high temperature expansion from some reference system whose properties are supposed to be
already known or exactly solved \cite{Zwanzig,Barker1}. 
In particular, it was shown that this approach can be even applicable to the lower temperature and high density regions.
However, this method is usually used to derive global thermodynamical quantities
and we need a new framework which enables us to calculate various local quantities
such as the correlation functions.
One of  useful approaches toward understanding such local quantities is the field theoretical method
proposed by J. Hubbard and P. Schofield \cite{Hubbard}. 
In this method, the grand canonical partition function is cleverly transformed
into  a path integral formulation of a scalar field theory.   
In the following, we first review the method \cite{Hubbard}, 
and then calculate the corrections to the correlation functions. 

Suppose that a reference system is described by a two-body potential $v_R(x)$
and then perturbed as
\aln{v(x)=v_R^{}(x)+\epsilon v_1^{}(x),
\label{eq:perturbative potential}
}
where 
$\epsilon$ is a small parameter and $v_1^{}(x)$ is an arbitrary potential.   
Note that, in the case of the scale-transformation discussed in the previous section, we take
$v_R^{}(x)=v(x)$ itself  and $v_1^{}(x)=x^\mu\partial_\mu^{}v(x)$. 

Under the shift of Eq. (\ref{eq:perturbative potential}), 
the partition function Eq. (\ref{eq: partition 2}) becomes 
\aln{
& \Xi_{v_R^{}+\epsilon v_1^{}}^{}[T,\mu,V;U(x)] \nn
&=\sum_{N=0}^\infty \frac{z(\mu+\epsilon v_1^{}(0)/2)^N}{N!}
\int_V 
\cdots
\int_V \prod_{i=1}^N d^dx_i^{} \ e^{-\beta \sum_{i<j}v_R^{}(x_i^{}-x_j^{})}
e^{
\left(
-\frac{\beta}{2}\langle \rho|\epsilon v_1^{}|\rho\rangle+\langle \beta U | \rho\rangle
\right) }
\nn
&=\Xi_{v_R^{}}^{}[T,\mu+\frac{\epsilon v_1^{}(0)}{2},V;0]\left\langle e^{-\frac{1}{2}\langle \rho|\beta \epsilon v_1^{}|\rho\rangle+\langle \beta U|\rho\rangle}\right\rangle_R^{},
\label{eq: perturbation1}
}
where $\langle \cdots \rangle_R^{}$ represents the thermodynamical expectation value 
in the reference system: 
\aln{\langle {\cal{O}}\rangle_R^{}\equiv \frac{1}{\Xi_{v_R^{}}^{}[T,\mu+\frac{\epsilon v_1^{}(0)}{2},V;0]}\sum_{N=0}^\infty \frac{z(\mu+\frac{\epsilon v_1^{}(0)}{2})^N}{N!}
\int_V 
\cdots
  \int_V  \prod_{i=1}^N d^dx_i^{} \ e^{-\beta \sum_{i<j}v_R^{}(x_i^{}-x_j^{})}
{\cal{O}}  .
} 
Here we have absorbed the constant $\epsilon v_1(0)$ into the the chemical potential
of the reference system, but instead we can simply set  $v_1(0)=0$ without changing any properties of liquids. 

In order to rewrite the partition function Eq. (\ref{eq: perturbation1}) 
in a  path integral form, we use the following mathematical identity:
\aln{
e^{\frac{a}{2}x^2}=\begin{cases}
\sqrt{2a\pi}\int_{-\infty}^\infty dy \ \exp\left(-\frac{y^2}{2a}+xy\right), & \text{for }a>0
\\
\sqrt{-2a\pi}\int_{-\infty}^\infty dy \ \exp\left(\frac{y^2}{2a}+ixy\right), & \text{for }a<0
\end{cases},
\label{eq:mathematical identity}
}
where  $a>0\ (<0)$  corresponds to an attractive (repulsive) potential respectively, 
i.e., $v_1^{}(x)<0\ (>0)$. 
In the following, we consider an attractive case for simplicity. 
\footnote{
In more general liquid/vapor systems such that there is some polarization dependences, we need more careful analysis because the coefficient $a$ can become complex in such a case.     
}

By completing the square in the exponent in Eq. (\ref{eq: perturbation1}), 
we obtain 
\aln{
&\exp\left(-\frac{1}{2}\langle \rho|\beta  \epsilon v_1^{}|\rho\rangle+\langle \beta U|\rho\rangle\right) \nn
&=\exp\left(-\frac{1}{2}\left\langle\rho-\frac{1}{\epsilon} Uv_1^{-1}\bigg|\beta  \epsilon v_1^{}\bigg|\rho-\frac{1}{\epsilon}v_1^{-1}U\right\rangle+\frac{1}{2}\left\langle \beta U\left|\frac{v_1^{-1}}{\beta \epsilon }\right|\beta U\right\rangle\right)
\nn
&={\cal{N}} e^{\frac{1}{2}\left\langle \beta U\left|\frac{v_1^{-1}}{\beta  \epsilon }\right|\beta U\right\rangle} 
\int{\cal{D}}\phi \exp\left(\frac{1}{2}\left\langle \phi\left|\frac{v_1^{-1}}{\beta  \epsilon }\right|\phi \right\rangle+\left\langle\rho-\frac{U}{ \epsilon }v_1^{-1}\bigg|\phi\right\rangle
\right),
}
where $v_1^{-1}$ is an inverse operator acting on functions, and in the second equality, 
we have used Eq. (\ref{eq:mathematical identity}) 
to rewrite the first term in the second line as a path integral over the new variable $\phi$.  
Eq. (\ref{eq: perturbation1}) is rewritten in terms of the scalar field path integral:
\aln{
\frac{\Xi_{v_R^{}+\epsilon v_1^{}}^{}[T,\mu,V;U(x)]}{\Xi_{v_R^{}}^{}[T,\mu+\epsilon v_1^{}(0)/2,V;0]}
&={\cal{N}}  e^{\frac{1}{2}\left\langle \beta U\left|\frac{v_1^{-1}}{\beta \epsilon}\right|\beta U\right\rangle}
\int{\cal{D}}\phi 
e^{ \frac{1}{2}\left\langle \phi\left|\frac{v_1^{-1}}{\beta \epsilon }\right|\phi \right\rangle-\left\langle\frac{U}{\epsilon }
\big|v_1^{-1}\big|\phi\right \rangle }
\langle e^{\langle\rho|\phi\rangle}\rangle_R^{},
}
where ${\cal{N}} $ is a normalization factor.
Then, the cumulant expansion of $\langle e^{\langle\rho|\phi\rangle}\rangle_R^{}$ leads to the following result:  
\aln{
  \frac{\Xi_{v_R^{}+\epsilon v_1^{}}^{}[T,\mu,V;U(x)]}{\Xi_{v_R^{}}^{}[T,\mu+\epsilon v_1^{}(0)/2,V;0]}
& ={\cal{N}}\int{\cal{D}}\phi 
e^{\frac{1}{2}\left\langle \phi\left|\frac{v_1^{-1}}{\beta \epsilon }\right|\phi \right\rangle-\left\langle\frac{U}{\epsilon }\big| v_1^{-1}\big |\phi\right\rangle }
\nn
& \times
\exp \bigg(
\sum_{l=1}^\infty \frac{1}{l!}
\int \prod_{i=1}^l \frac{d^dk_i^{}}{(2\pi)^d} \tilde{F}_l^{(R)}(k_1^{},\cdots,k_l^{})\tilde{\phi}(k_1^{})\cdots \tilde{\phi}(k_l^{})
\bigg),
\label{eq: grand canonical by Hubbard}
}
where $\tilde{F}_l^{(R)}(k_1^{},\cdots,k_l^{})$ is the Fourier transform of the correlation function of the reference system. 
This result shows that a classical theory of liquid is equivalent to a quantum field theory 
with an infinite number of the multiple point operators,  each of which corresponds to the correlation function of the reference system. 

In this expression, we see that the external source $U(x)$ originally introduced for the density $\rho(x)$ now plays a role of the source term for the quantum field $\phi(x)$. 
Therefore $\phi(x)$ can be essentially identified with $\rho(x)$. 
\footnote{There is an extra factor $v_1^{-1}/\epsilon$ in the
quadratic and a linear term of $\phi$ in the Lagrangian of Eq. (\ref{eq: grand canonical by Hubbard}) but
these factors are cancelled in $\epsilon =0$ limit. Thus $U$ derivatives give the correlation
functions of $\phi(x)$.}
In particular, it is apparent that the correlation function $F_l^{}(x_1^{},\cdots,x_l^{})$ of the density fluctuations
corresponds to the connected part of $\langle \phi(x_1^{})\cdots\phi(x_l^{})\rangle$ 
because it is generated by the generating functional $-\beta W[U]=\log \Xi[U]$.  
Therefore, we can calculate the perturbative corrections of these correlation functions in the same manner as 
the ordinary quantum field theories.  
Note that this procedure is widely applicable to any perturbed (or transformed) system as long as the resultant potential is given by  Eq. (\ref{eq:perturbative potential}). 

\vspace{5mm}
Let us now apply this method to our scale-transformed system, i.e. $v_R^{}(x)=v(x),\ v_1^{}(x)= \delta v(x)$. 
In this case, 
the correlation functions of the reference system $\tilde{F}_l^{(R)}(k_1^{},\cdots,k_l^{})$ is 
given by the exact correlation function $\tilde{F}_l^{}(k_1^{},\cdots,k_l^{})$ at some fixed density. 
They are, of course, not yet known. 
Instead, we investigate how they change under a scale transformation, 
namely under a change of the density. 

In calculating corrections to the correlators $\Delta \tilde{F}_l$, 
it is convenient to use Feynman diagrammatic representations. 
Associated with the propagator and $l$-point vertices, 
we introduce the following graphical representations:
\begin{itemize}
\item For each internal propagator for $\phi(x)$,   
\aln{\begin{minipage}{4cm}
\includegraphics[width=4cm]{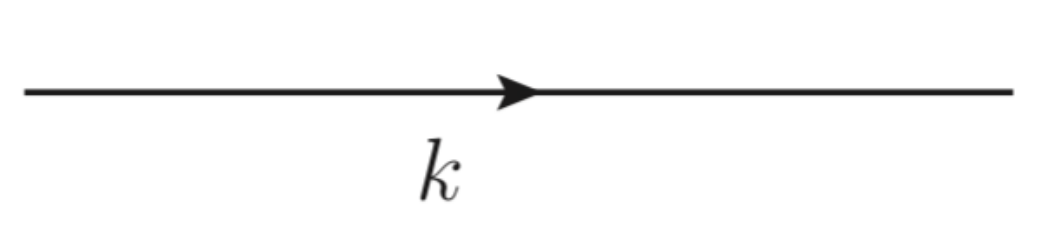}
\end{minipage}
=-\int \frac{d^dk}{(2\pi)^d}\beta\epsilon \delta\tilde{v}(k).
}
\item For each $l$-th vertex,
\aln{\begin{minipage}{3cm}
\includegraphics[width=3cm]{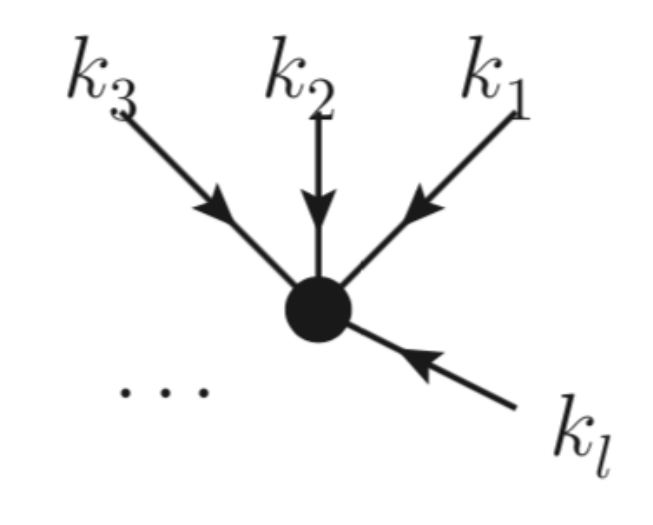}
\end{minipage}
=\tilde{F}_l^{}(k_1^{},\cdots,k_l^{}), 
}
\end{itemize}
where $\tilde{F}_l (k_1, \cdots, k_l)$ is symmetric under permutations of momenta and vanishes unless momentum conservation $\sum_{i=1}^l k_i$ is satisfied. 

The Fourier transform of $\delta v(x)$ is given by 
\aln{\delta\tilde{v}(k)=\int d^dxx^\mu\partial_{\mu}^{}v(x)e^{-ikx}=-\frac{\partial}{\partial k_\mu^{}}(k_\mu^{}\tilde{v}(k))=-d\tilde{v}(k)+k_\mu^{}\frac{\partial}{\partial k_\mu^{}}\tilde{v}(k).
}
Thus  we have $\delta \tilde{v}(0) = - d \tilde{v}(0)$.

In the graphical representation, there is an important property derived from  Eq. (\ref{eq: grand canonical by Hubbard}). 
When we take the functional derivatives of Eq. (\ref{eq: grand canonical by Hubbard}) with respect to $\beta\tilde{U}(p)$, an additional factor $-(\epsilon\delta \tilde{v}(p))^{-1}$ is added to each of the external legs. 
However, this additional factor is completely canceled by the propagator $-\epsilon \delta \tilde{v}(p)$. 
This property is depicted as 
\begin{itemize} 
\item For each external line, 
\aln{\begin{minipage}{8cm}
\includegraphics[width=8cm]{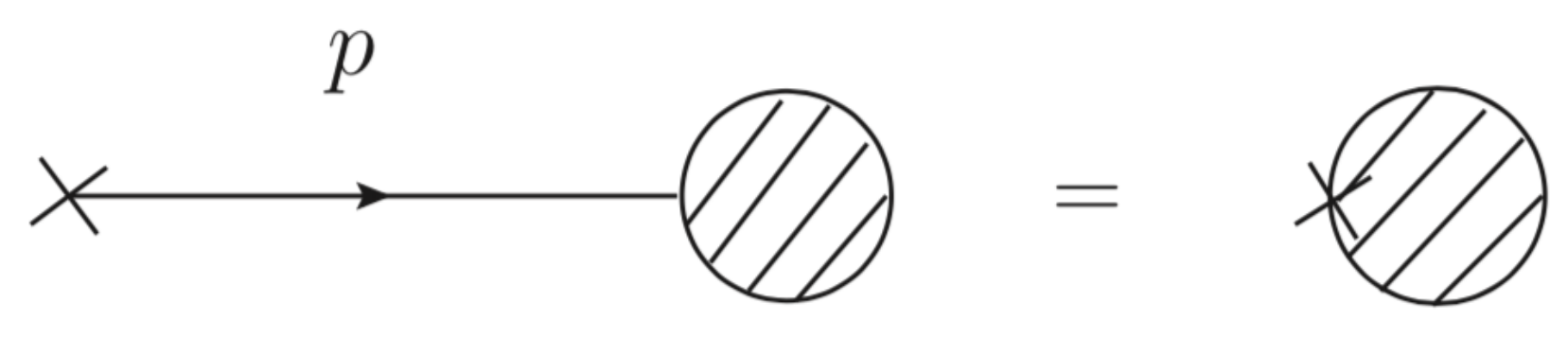}
\end{minipage},
}
\end{itemize}
where the blob is any diagram connected with this external line. 
As a result, the tree-level correlation functions $\tilde{F}_l^{}(p_1^{},\cdots,p_l^{})$, namely correlation functions without perturbation $\delta v$, can be correctly reproduced.

By using them, we can diagrammatically evaluate the perturbative corrections of the correlation functions, i.e. $\Delta F_l^{}(x_1^{},\cdots,x_l^{})$ in Eqs. (\ref{eq: PDE1}), (\ref{eq: PDE2}), and (\ref{eq: PDE3}).
Because we are interested in the $\epsilon\rightarrow 0$ limit, it is sufficient to consider the leading order contributions with respect to $\epsilon$. 
Such contributions are represented by the diagrams which contain only one internal propagator because it is the only place where an additional $\epsilon$ factor appears.  
In other words, if there are more than one internal propagators, the diagram vanishes in the
$\epsilon \rightarrow 0$ limit. 
Therefore, we obtain the following results:  
\\
\\
(0) {\bf Zero point function (=Grand potential):}
\aln{\Delta F_0^{} &=-\beta \delta W[T,\mu,V]=
\begin{minipage}{6cm}
\includegraphics[width=6cm]{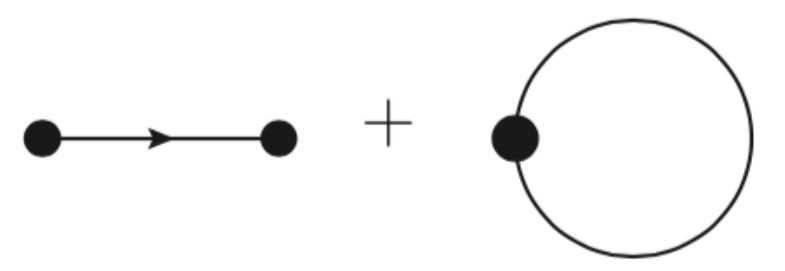}
\end{minipage}
\nn
&=-\frac{1}{2!}\int \frac{d^dp}{(2\pi)^d}\beta\delta\tilde{v}(p)\tilde{F}_1^{}(p)\tilde{F}_1^{}(-p)
-\frac{1}{2!}\int \frac{d^dp}{(2\pi)^d}\beta\delta\tilde{v}(p)\tilde{F}_2^{}(p,-p).
\label{eq: correction of W}
}
\\
$(1)$ {\bf One-point function:}
\aln{\Delta \tilde{F}_1^{}(k)&=
\begin{minipage}{8cm}
\includegraphics[width=8cm]{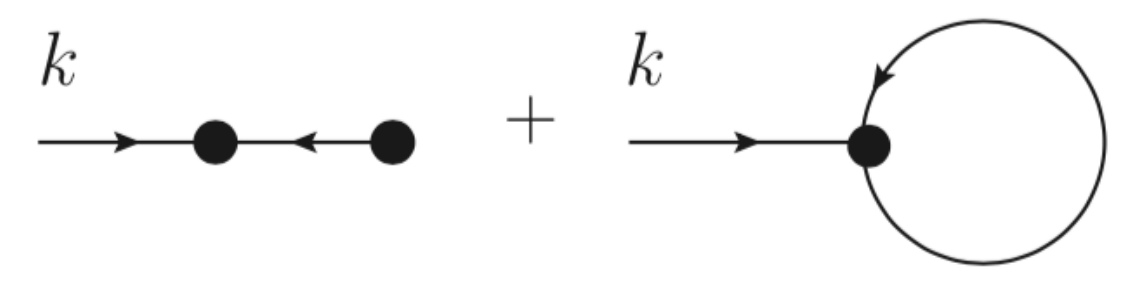}
\end{minipage}
\nn
&=-\frac{2!}{2!}\int \frac{d^dp}{(2\pi)^d}\beta\delta\tilde{v}(p)\tilde{F}_1^{}(p)\tilde{F}_2^{}(-p,k)
-\frac{3}{3!}\int \frac{d^dp}{(2\pi)^d}\beta\delta\tilde{v}(p)\tilde{F}_3^{}(p,-p,k).
}
\\
$(2)$ {\bf Two-point correlation function:}
\aln{\Delta \tilde{F}_2^{}(k_1^{},k_2^{})&=
\begin{minipage}{10cm}
\includegraphics[width=10cm]{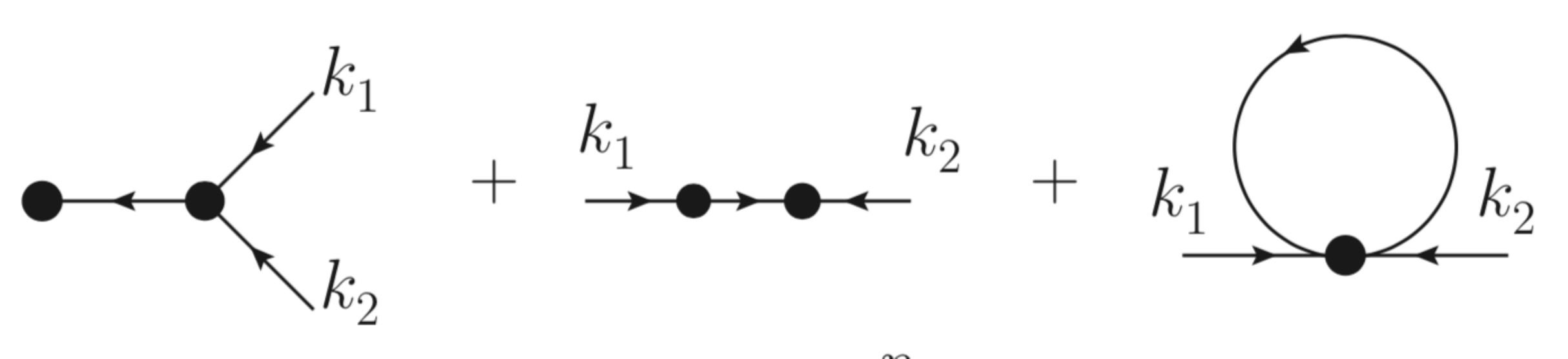}
\end{minipage}
\nn
=&-\frac{3!}{3!}\int \frac{d^dp}{(2\pi)^d}\beta\delta\tilde{v}(p)\tilde{F}_1^{}(p)\tilde{F}_3^{}(-p,k_1^{},k_2^{})
-\frac{2!2!2!}{(2!)^3}\int \frac{d^dp}{(2\pi)^d}\beta\delta\tilde{v}(p)\tilde{F}_2^{}(k_1^{},-p)\tilde{F}_2^{}(p,k_2^{})
\nn
&-\frac{4\cdot 3}{4!}\int \frac{d^dp}{(2\pi)^d}\beta\delta\tilde{v}(p)\tilde{F}_4^{}(k_1^{},k_2^{},p,-p).
}
\\
$(3)$ {\bf Three-point correlation function:}
\aln{
\Delta & \tilde{F}_3^{}(k_1^{},k_2^{},k_3^{})=
\begin{minipage}{10cm}
\includegraphics[width=10cm]{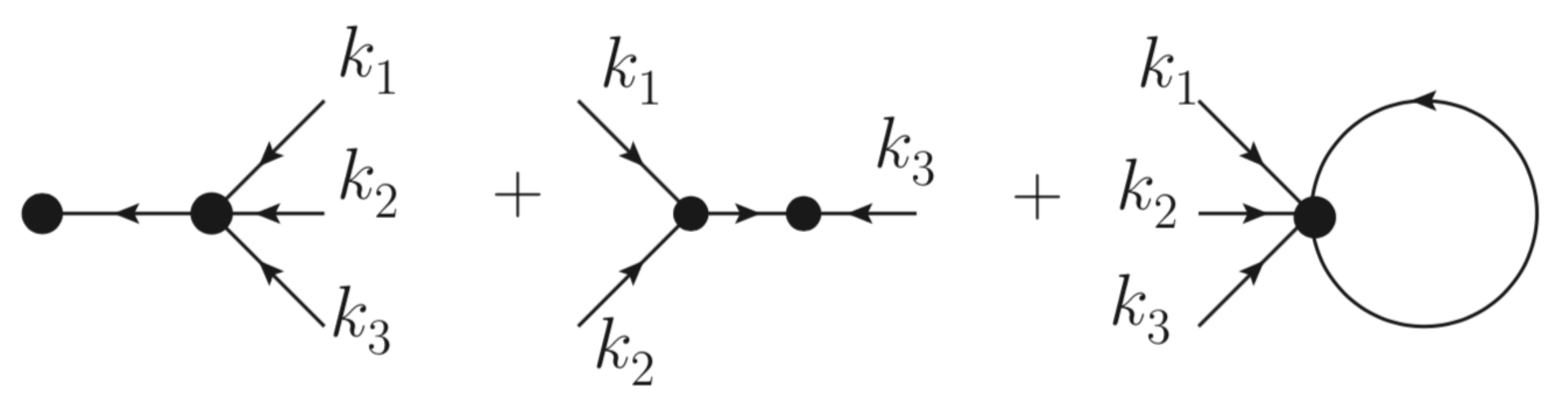}
\end{minipage}
\nn
=&-\frac{4!}{4!}\int \frac{d^dp}{(2\pi)^d}\beta\delta\tilde{v}(p)\tilde{F}_1^{}(p)\tilde{F}_4^{}(-p,k_1^{},k_2^{},k_3^{})
\nn
&-\frac{3!\cdot 2!}{3!2!}\int \frac{d^dp}{(2\pi)^d}\beta\delta\tilde{v}(p)\left[\tilde{F}_3^{}(k_1^{},k_2^{},-p)\tilde{F}_2^{}(p,k_3^{})+\tilde{F}_3^{}(k_2^{},k_3^{},-p)\tilde{F}_2^{}(p,k_1^{})+\tilde{F}_3^{}(k_3^{},k_1^{},-p)\tilde{F}_2^{}(p,k_2^{})\right]
\nn
&-\frac{5\cdot 4\cdot 3}{5!}\int \frac{d^dp}{(2\pi)^d}\beta\delta\tilde{v}(p)\tilde{F}_5^{}(k_1^{},k_2^{},k_3^{},p,-p).
}
\\
\\
$(4)$ {\bf $l$-point correlation functions $(l\geq 4)$:}\\
In general, $\Delta \tilde{F}_{l+1}^{}(k_1^{},\cdots,k_{l+1}^{})$ can be automatically obtained by taking the functional derivative of the $l$-th diagrams and using the following vertex relation:
\aln{\frac{\delta \tilde{F}_l^{}(k_1^{},\cdots,k_l^{})}{\delta \beta\tilde{U}(k_{l+1}^{})}\bigg|_{U=0}^{}=\tilde{F}_{l+1}^{}(k_1^{},\cdots,k_{l+1}^{}),
}
which is diagrammatically represented by
\aln{\frac{\delta}{\delta (\beta \tilde{U}(k_{l+1}^{}))}
\begin{minipage}{6cm}
\includegraphics[width=6cm]{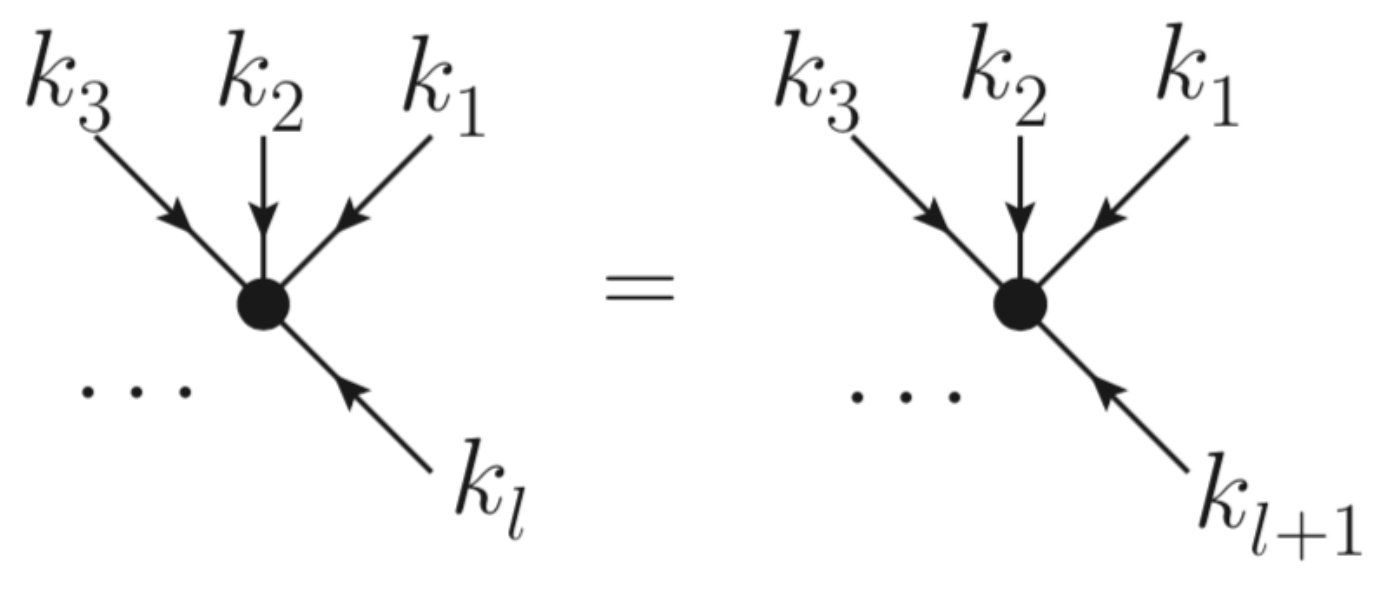}
\end{minipage}.
}
%
In fact, we can straightforwardly check that all the above results  can be 
derived by taking the functional derivatives of $\Delta F_0^{}=-\beta\Delta W$. 
%
\subsection{DRGEs as hierarchical equations}
\label{subsec:Hirarchical}
By substituting the results of the previous section into Eqs. (\ref{eq: PDE1}), (\ref{eq: PDE2}), and (\ref{eq: PDE3}), 
we obtain a sequence of differential equations that govern the changes of the correlation functions against a
variation of the density (or chemical potential).
In order to separate the momentum conservation from the correlation functions, 
we introduce the following notations:
\aln{-\frac{\beta W}{nV}&=\frac{p}{nT}\equiv\lambda,
\label{eq: coupling 1}
\\
\tilde{F}_1^{}(k)&=(2\pi)^d\delta^{(d)}(k)n,
\label{eq: coupling 2}
\\
 \tilde{F}_2^{}(k_1^{},k_2^{})&=(2\pi)^d\delta^{(d)}(k_1^{}+k_2^{})\tilde{\kappa}(k_1^{},k_2^{})
=(2\pi)^d\delta^{(d)}(k_1^{}+k_2^{})n\kappa(k_1^{},k_2^{}) .
\label{eq: coupling 3}
}
Especially, the 2-point function at zero momentum $\kappa(k=0)$ is identified with $\kappa_T.$ 
For higher $l \ge 3$, we define
\aln{
\tilde{F}_l^{}(k_1^{},\cdots,k_l^{})&=(2\pi)^d
\delta^{(d)}\left(\sum_{i=1}^l k_i \right)
\tilde{\lambda}_l^{}(k_1^{},\cdots,k_l^{})
=(2\pi)^d
\delta^{(d)}\left(\sum_{i=1}^l k_i \right)
n\lambda_l^{}(k_1^{},\cdots,k_l^{}) .
\label{eq: coupling 4}
}
Note that 
$\lambda$, $ \kappa$ and $\lambda_l^{}\ (l\geq 3)$ are dimensionless. 
Especially, 
their initial values at $n=0\ (\mu=-\infty)$, namely the values for the ideal gas,
 are given by
\aln{\lambda|_{n=0}^{}=1,\ \kappa(k_1^{},k_2^{})|_{n=0}^{}=1,\ 
\lambda_l^{}(k_1^{},\cdots,k_l^{})|_{n=0}^{}=1\ (l\geq 3),
}
as can be easily checked from the correlation functions for the ideal gas (see \ref{app:ideal}). 

\vspace{10mm}
By { substituting Eqs. (\ref{eq: coupling 1}), (\ref{eq: coupling 2}), (\ref{eq: coupling 3}), and 
(\ref{eq: coupling 4}) into Eq. (\ref{eq: PDE2}) or Eq. (\ref{eq: PDE3}), 
we obtain the following set of hierarchical equations.}\footnote{
The momentum derivative in the LHS of Eq.(\ref{eq: PDE2}) or (\ref{eq: PDE3}) is given by
\aln{
-\sum_{i=1}^lk_i^\mu & \partial_{i\mu}^{}\left(\delta^{(d)}\left(\sum_{i=1}^lk_i^{}\right) \tilde{\lambda}_l^{}\right)
=- \tilde{\lambda}_l^{}\sum_{i=1}^l k_i^\mu \partial_{i\mu}^{}\delta^{(d)}\left(\sum_{i=1}^lk_i^{}\right)
-\sum_{i=1}^l k_i^\mu \delta^{(d)}\left(\sum_{i=1}^lk_i^{}\right)\partial_{i\mu}^{}\tilde{\lambda}_l^{}
\nn
&=\delta^{(d)}\left(\sum_{i=1}^lk_i^{}\right)\left(d\tilde{\lambda}_l^{}-\sum_{i=1}^lk_i^{\mu}\partial_{i\mu}^{}\tilde{\lambda}_l^{} \right).
}
}
For $l=0$, we get an equation of  state of the classical liquid:
\aln{&\frac{p}{T}-n
=\frac{1}{2}\beta\tilde{v}(0)n^2-\frac{1}{2d}\int\frac{d^dp}{(2\pi)^d} \beta\delta\tilde{v}(p) \tilde{\kappa}(p,-p),
\label{eq: rge for potential}
}
which relates the pressure with an integral of two-point correlation function. 
Here we have used $ \delta \tilde{v}(0)=-d\tilde{v}(0).$

For $l \ge 1$ we have the following set of partial differential equations. First we define
the differential operator ${\cal D}$ as
\aln{
{\cal{D}}&\equiv d\left(-\frac{\partial}{\partial(\beta \mu)}\bigg|_{V,T}^{}+\frac{\partial}{\partial\log V}\bigg|_{\mu,T}^{}+1-\frac{1}{d}\sum_{i=1}^lk_i^\mu\frac{\partial}{\partial k_i^\mu}
\right)
\\
&=d\left(-\kappa_T^{}\frac{\partial}{\partial \log n}\bigg|_{V,T}^{}+\frac{\partial}{\partial\log V}\bigg|_{\mu,T}^{}+1
-\frac{1}{d}\sum_{i=1}^lk_i^\mu\frac{\partial}{\partial k_i^\mu}
\right) .
}
Here we note that the volume derivative in the  differential operator ${\cal D}$ can be neglected
in the large $V$ limit since local quantities such as the density $n$ or the correlation functions 
do not depend on the total volume when $(\mu, T)$ are fixed. 

For $l=1$, we get 
\aln{
&
{\cal{D}}n(k)
=d\beta\tilde{v}(0)n\tilde{\kappa}(k,0)-\frac{1}{2}\int\frac{d^dp}{(2\pi)^d} \beta\delta\tilde{v}(p) \tilde{\lambda}_3^{}(p,-p,k).
\label{eq:rge2}
}
Because of the assumption of the translational invariance, it vanishes unless $k =0.$ 
Since $\mu$ derivative in the left hand side for $k=0$ is written in terms of  $\kappa_T=\kappa(0)$,
the equation relates $\kappa_T$ with an integral of the 3-point function $\lambda_3$.

For $l =2$, we have
\aln{
{\cal{D}}\tilde{\kappa}(k_1^{},k_2^{})
=& d\beta\tilde{v}(0)n\tilde{\lambda}_3^{}(k_1^{},k_2^{},0)
-\frac{1}{2}\tilde{\kappa}(k_1^{},k_2^{})
\sum_{i=1}^2 
\beta\delta \tilde{v}(k_i^{})\tilde{\kappa}(k_i^{},-k_i^{})
\nn
& 
-\frac{1}{2}\int\frac{d^dp}{(2\pi)^d} 
\beta\delta\tilde{v}(p) \tilde{\lambda}_4^{}(k_1^{},k_2^{},p,-p),
\label{eq:rge3}
}
which relates a density response of $\kappa(k)$ with an integral of 3- and 4-point functions.
For $l =3$, we have
\aln{
&
{\cal{D}}\tilde{\lambda}_3^{}(k_1^{},k_2^{},k_3^{})
=d\beta\tilde{v}(0)n \tilde{\lambda}_4^{}(k_1^{},k_2^{},k_3^{},0)
-\tilde{\lambda}_3^{}(k_1^{},k_2^{},k_3^{})
\sum_{i=1}^3 
\beta\delta\tilde{v}(k_i^{})\kappa(k_i^{},-k_i^{})
\nn
&\h{3cm} 
-\frac{1}{2}\int\frac{d^dp}{(2\pi)^d} \beta\delta\tilde{v}(p) 
\tilde{\lambda}_5^{}(k_1^{},k_2^{},k_3^{},p,-p) .
}
For $l=4$, see \ref{app-F4}. 

These hierarchical equations  describe the response of the system to a small change of the density. 
Though we have used perturbative technique, 
they are the exact (non-perturbative) equations 
and offer us an alternative formulation of the classical liquid/vapor system. 
Our next step is to solve them by using physically reasonable approximations or assumptions. 
In \ref{app-solve}, we will briefly discuss how we can attack to solve the set of differential equations (DRGE)
derived in the previous section, but as we will see in the next section, it will be much better to first
make the Legendre transformations to  one-particle irreducible (1PI) diagrams 
in which multipoint 1PI vertices are expected to become local and 
 closures of the hierarchical equations become more reliable.


\section{Legendre Transformation and 1PI Potential
} \label{sec:Helmholtz}
In quantum field theory, it is usually much more convenient 
to discuss the dynamics of a system based on the effective action $\Gamma[\phi]$ which is obtained by the Legendre transformation of  $iW[J]=\log Z[J]$; $\Gamma[\phi]$ represents the generating functional of the 1PI diagrams.  
The 1PI effective action $\Gamma[\phi]$ is especially useful and inevitable
when a spontaneous symmetry breaking occurs. 
Similarly 2PI effective action is important when we discuss a nontrivial behavior of the propagator \cite{Berges:2004yj}. 
In this paper, we concentrate on $\Gamma[\phi]$ 
and leave analysis of 2PI actions for future.

In the case of the classically interacting particles, such a transformation corresponds to the thermodynamical Legendre transformation of the grand potential $-\beta W[U]$ to the Helmholtz free energy $-\beta \Gamma[\rho]$ where the thermodynamical parameters are transformed from $(T,V,\mu)$ to $(T,V,N)$.  
Either thermodynamical potential has its advantage and we can use them as the situation demands.
Here we generalize the thermodynamical Legendre transformation including the local 
external source term $U(x)$. Thus, $-\beta W_v^{}[U]$ is transformed to a generating functional
of 1PI correlations functions of density, i.e. $-\beta\Gamma[\rho]$. 
Up to trivial contributions from the ideal gas, 
such 1PI correlation functions are called the
 {\it direct correlation functions} in the liquid theory. 

In the following, in order to make the discussions simpler, we absorb
the chemical potential $\mu$ into the zero mode of the external source $U(x)$
and denote $-\beta W_v^{}[T,\mu,V;U]$ as $-\beta W_v^{}[T,V;U]$.
Also we introduce the correlation functions of density fluctuations in the presence of
external source terms and denote them as
\aln{F_l^U(x_1^{},\cdots,x_l^{})\equiv \frac{\delta^l(-\beta W_v^{}[U])}{\delta(\beta U(x_1^{}))\cdots\delta(\beta U(x_l^{}))} .
}
Setting $U(x)=\mu$, they 
coincide with the previous ones $F_l^{}(x_1^{},\cdots,x_l^{})$. 
%

\subsection{
Direct correlation functions}    
We define the (generalized) Helmholtz free energy $-\beta\Gamma_v^{}[T,V;\rho]$ 
by the Legendre transformation of $-\beta W_v^{}[T,V;U]$: 
\aln{-\beta\Gamma_v^{}[T,V;\rho]&=\underset{U}{\text{Min}}\left(-\beta W_v^{}[T,V;U]-\beta \int d^dxU(x)\rho(x)\right)
\nn
&=-\beta W_v^{}[T,V;U_\rho^{v}]-\beta \int d^dxU_\rho^{v}(x)\rho(x),
}
where $\rho(x)$ represents a density field, and $U_\rho^{v}(x)$ is a solution of 
\aln{\frac{\delta(-\beta W_v^{}[T,V;U])}{\delta (\beta U(x))}=\rho(x).
\label{eq: def of minimum}
}
Then, we can define new correlation functions by taking the functional derivatives of $-\beta\Gamma_v^{}[T,V;\rho]$ with respect to $\rho(x)$:
\aln{c_l^{}(x_1^{},x_2^{},\cdots,x_l^{})\equiv \frac{\delta^l(-\beta \Gamma_v^{}[T,V;\rho])}{\delta \rho(x_1^{})\delta \rho(x_2^{})\cdots \delta \rho(x_l^{})}\bigg|_{\rho(x)=n}^{}.
}
In the following, we call them the $l-$point 1PI vertices for $l \ge 3$. 
In particular, $c_2^{}(x,y)$ and $c_3^{}(x,y,z)$ satisfy the following relations
\aln{
&\int d^dzF_2^{}(x,z)c_2^{}(z,y)=-\delta^{(d)}(x-y),
\label{eq:inverse relation}
\\
&F_3^{}(x,y,z)=\int d^dw\int d^dw'\int d^dw''F_2^{}(x,w)F_2^{}(y,w')F_2^{}(z,w'')c_3^{}(w,w',w''),
\label{eq:three point relation}
}
where $F_2$ and $F_3$ are the (connected) correlation functions defined in the previous section. 
These relations are the direct consequences of the Legendre transformation. 
The first relation is called the ``{\it Ornstein-Zernike equation}" in  the liquid/vapor theory.  
In order to rewrite it in a standard form, we note
that the two-point ``direct correlation function" $c_2^D(x,y)$ is  defined by  
\aln{ c_2^D(x,y)=c_2^{}(x,y)+n^{-1}\delta^{(d)}(x-y).}  
The direct correlation function $c_2^D$ vanishes for the ideal gas (see \ref{app:ideal}). 
Then using the definition of the total correlation function
$h_2$ in Eq. (\ref{eq:total correlation function}) which also vanishes for the ideal gas,   
Eq. (\ref{eq:inverse relation}) becomes
\aln
{h_2^{}(x,y)=c_2^D(x,y)+n\int d^dyc_2^D(z,y)h_2^{}(x,z) , 
\label{eq:OZ}
}
which is the standard form of the Orstein-Zernike equation.

Similarly we define $l$-point direct 1PI vertices (for $l \ge 3$), $c_l^D$, by
\aln{
c_l^D(x_1, \cdots , x_l)=c_l^{}(x_1, \cdots , x_l)-\frac{(-1)^{l+1}(l-2)!}{n^{l-1}}\prod_{i=2}^l \delta^{(d)}(x_1-x_i).
}
The direct 1PI vertices are defined so as to vanish for the ideal gas (see \ref{app:ideal}).

\subsection{Scale transformations of 1PI potential}
Now let us consider a scale transformation of the potential
 $v(x)\rightarrow v(x)+\epsilon\delta v(x)$ and its consequence to the Helmholtz free energy
  $-\beta\Gamma_{v}^{}[T,V;\rho]$. 
Here, we should recall the relation Eq. (\ref{eq:relation1})
between $-\beta W_{v+\epsilon \delta v}^{}[T,V;U]$ and $-\beta W_v^{}[T,V;U]$:
\aln{
-\beta W_{v+\epsilon \delta v}^{}[T,V;U(x)]
=-\beta W_v^{}[T,(1+d\epsilon)V;U(x(1-\epsilon))-dT\epsilon].
}
The Legendre transformation of the LHS gives $-\beta \Gamma_{v+\epsilon \delta v}^{}[T,V;\rho(x)]$ by  definition. 
On the other hand, the Legendre transformation of the RHS is
\aln{
&\underset{U}{\text{Min}}\left[-\beta W_v^{}[T,(1+d\epsilon)V;U(x(1-\epsilon))-dT\epsilon]-\beta\int_{V}d^dxU(x)\rho(x)
\right] 
\nn
=& \underset{U}{\text{Min}}\bigg[-\beta W_v^{}[T,(1+d\epsilon)V;U(x(1-\epsilon))-dT\epsilon]  \nn
& -(1-d\epsilon)\beta\int_{(1+d\epsilon)V}d^dx(U(x(1-\epsilon))-dT\epsilon)\rho(x(1-\epsilon)) 
\bigg]
-d\epsilon\int_V d^dx\rho(x)
\nn
=&-\beta \Gamma_v^{}[T,(1+d\epsilon)V;(1-d\epsilon)\rho(x(1-\epsilon))]-d\epsilon\int_V^{} d^dx\rho(x),
}
where we have regarded $\tilde{U}(x)=U(x(1-\epsilon))-dT\epsilon$ as a new external source and performed the Legendre transformation with respect to it.    
Thus, we obtain 
\aln{-\beta \Gamma_{v+\epsilon \delta v}^{}[T,V;\rho(x)]=-\beta \Gamma_v^{}[T,(1+d\epsilon)V;(1-d\epsilon)\rho(x(1-\epsilon))]-d\epsilon\int_V^{} d^dx\rho(x)+{\cal{O}}(\epsilon^2),
}
and, 
by differentiating it with respect to $\rho(x)$, 
we obtain the following differential equation of the 1PI vertices:
\aln{
\left(d\frac{\partial}{\partial\ln V}\bigg|_{T,N}^{}+\sum_{i=1}^lx_i^\mu\partial_{i\mu}^{}\right) c_l^{}(x_1^{},\cdots,x_l^{})-d\delta_{l0}^{}N-\delta_{l1}^{}d
=\Delta c_l^{}(x_1^{},\cdots,x_l^{}),
\label{eq: PDF for direct correlation}
}
where $\Delta c_l^{}(x_1^{},\cdots,x_l^{})$ represents the perturbative corrections caused by $\delta v(x)$. 
They are explicitly calculated in the next section. 
Note that the $V$ derivative in the LHS is equivalent to the $n(=\langle N\rangle /V)$ derivative 
\aln{
\frac{\partial}{\partial\ln V} = - \frac{\partial}{\partial \ln n}
}
because $\langle N \rangle$ is fixed here.
Thus the equation describes a response of various 1PI quantities 
against a small change of density. 
For the ideal gas, the LHS is shown to vanish (see \ref{app:ideal}). 

\subsection{Derivations of $\Delta c_l^{}(x_1^{},\cdots,x_l^{})$}
In this section, we will explicitly calculate the corrections $\Delta c_l^{}(x_1^{},\cdots,x_l^{})$.
%
In order for this, we use the previous result of the correction to the grand potential;  
\aln{-\beta W_{v+\epsilon\delta v}^{}[T,V;U]=-\beta W_{v}^{}[T,V;U]-\beta \Delta W_v^{}[T,V;U],
}
where $-\beta \Delta W_v^{}[T,V;U]$ is given by Eq. (\ref{eq: correction of W}). 
Then, from the definition of the Legendre transformation, we have
\aln{
& -\beta \Gamma_{v+\epsilon \delta v}[T,V;\rho] = 
\underset{U(x)}{\text{Min}}\left[-\beta W_{v}^{}[T,V;U]-\beta\Delta W_v^{}[T,V;U]+{\cal{O}}(\epsilon^2)-\beta \int d^dxU(x)\rho(x)\right]
\nn
& \hspace{10mm}  =-\beta W_v^{}[T,V;U_\rho^{v+\epsilon\delta v}]-\beta\Delta W_v^{}[T,V;U_\rho^{v+\epsilon\delta v}]-\beta \int d^dxU_\rho^{v+\epsilon\delta v}(x)\rho(x)+{\cal{O}}(\epsilon^2),
}
where $U_\rho^{v+\epsilon\delta v}(x)$ is a solution of 
\aln{ \frac{\delta (-\beta W_v^{}[T,V;U]-\beta\Delta W_v^{}[T,V;U])}{\delta (\beta U(x))}=\rho(x).
}
Thus, by denoting the difference between $U_\rho^{v+\epsilon\delta v}(x)$ and $U_\rho^{v}(x)$ as $\Delta U_\rho^v(x)$, we have
\aln{
-\beta \Gamma_{v+\epsilon\delta v}[T,V;\rho]& =-\beta W_v^{}[T,V;U_\rho^{v}]+\int d^dx \frac{\delta (-\beta W_v^{}[T,V;U])}{\delta (\beta U(x))}\bigg|_{U=U_\rho^v}^{}\beta\Delta U_\rho^{v}(x)
\nn
& \hspace{4mm} -\beta\Delta W_v^{}[T,V;U_\rho^v] 
-\beta \int d^dx(U_\rho^v(x)+\Delta U_\rho^{v}(x))\rho(x)+{\cal{O}}(\epsilon^2)
\nn
&= -\beta\Gamma_v^{}[T,V;\rho]-\beta\Delta W_v^{}[T,V;U_\rho^v]+{\cal{O}}(\epsilon^2),
\label{eq:correction of Gamma}
}
where we used Eq. (\ref{eq: def of minimum}).  
From Eq. (\ref{eq: correction of W}), the second term $-\beta\Delta W_v^{}[T,V;U_\rho^v]$ is given by 
\footnote{Here, $F_1^{U_\rho^v}(x)$ in Eq. (\ref{eq: correction of W}) becomes $\rho(x)$ by the definition;
\aln{F_1^{U_\rho^v}(x)=\frac{\delta (-\beta W_v^{}[T,V;U])}{\delta (\beta U(x))}\bigg|_{U=U_\rho^v}=\rho(x).
}
}
\aln{-\beta \Delta W_v^{}[T,V;U_\rho^v]&=-\frac{1}{2!}\int d^dx\int d^dy\rho(x)\beta\delta v(x-y)\rho(y)
-\frac{1}{2!}\int d^dx \int d^dy \beta\delta v(x-y)F_2^{U_\rho^v}(x,y).
}
Therefore, by functionally differentiating Eq.  with respect to $\rho(x)$ and putting $\rho(x)=n$, we can obtain $\Delta c_l^{}(x_1^{},\cdots,x_l^{})$. 
In order for systematic calculations, we introduce the following  graphical representations: 
\begin{itemize}
\item For the perturbative potential, we use the wavy line:
\aln{\beta \delta v(x-y)=
\begin{minipage}{5cm}
\includegraphics[width=5cm]{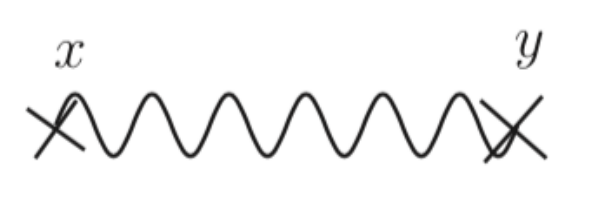}
\end{minipage}
\label{eq:potential propagator}
}
\item For the two-point (exact) correlation function, a straight line with a blob is used:
\aln{F_2^{U_\rho^v}(x,y)=
\begin{minipage}{5cm}
\includegraphics[width=5cm]{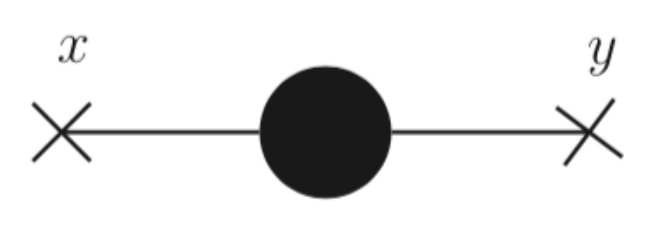}
\end{minipage}
}
\item For the 1PI vertices, shaded polygons are used:
\aln{c_l^{}(x_1^{},x_2^{},\cdots,x_l^{})=
\begin{minipage}{4cm}
\includegraphics[width=4cm]{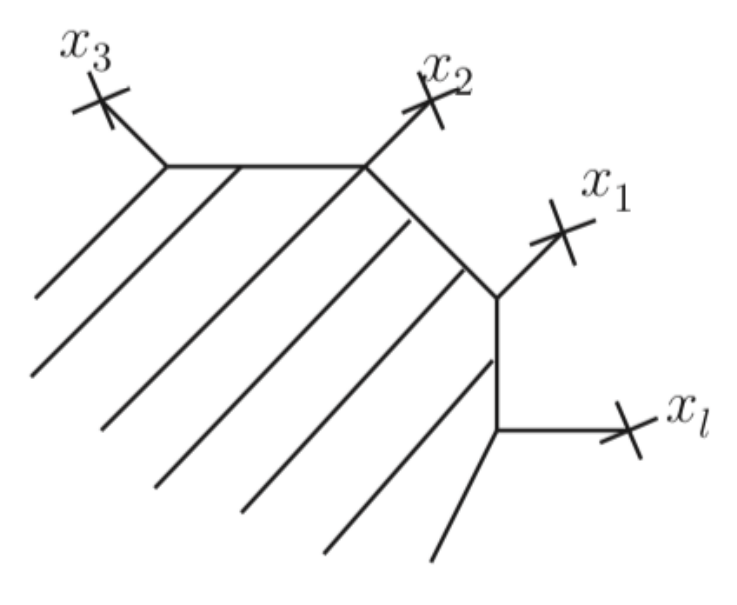}
\end{minipage}
}
\end{itemize}
These graphical representations are used to express $\Delta c_l$ in the following.

\vspace{5mm}
Here we note that 
the functional derivative of  $c_l^{}(x_1^{},\cdots,x_l^{})$ with respect to the density field $\rho$ is given by
\aln{\frac{\delta c_l^{}(x_1^{},\cdots,x_l^{})}{\delta\rho(x_{l+1}^{})}\bigg|_{\rho(x)=n}^{}=c_{l+1}^{}(x_1^{},\cdots,x_{l+1}^{}), 
\label{eq: density derivative} 
}
which is graphically represented by
\aln{
\frac{\delta}{\delta \rho(x_{l+1}^{})}
\begin{minipage}{8cm}
\includegraphics[width=8cm]{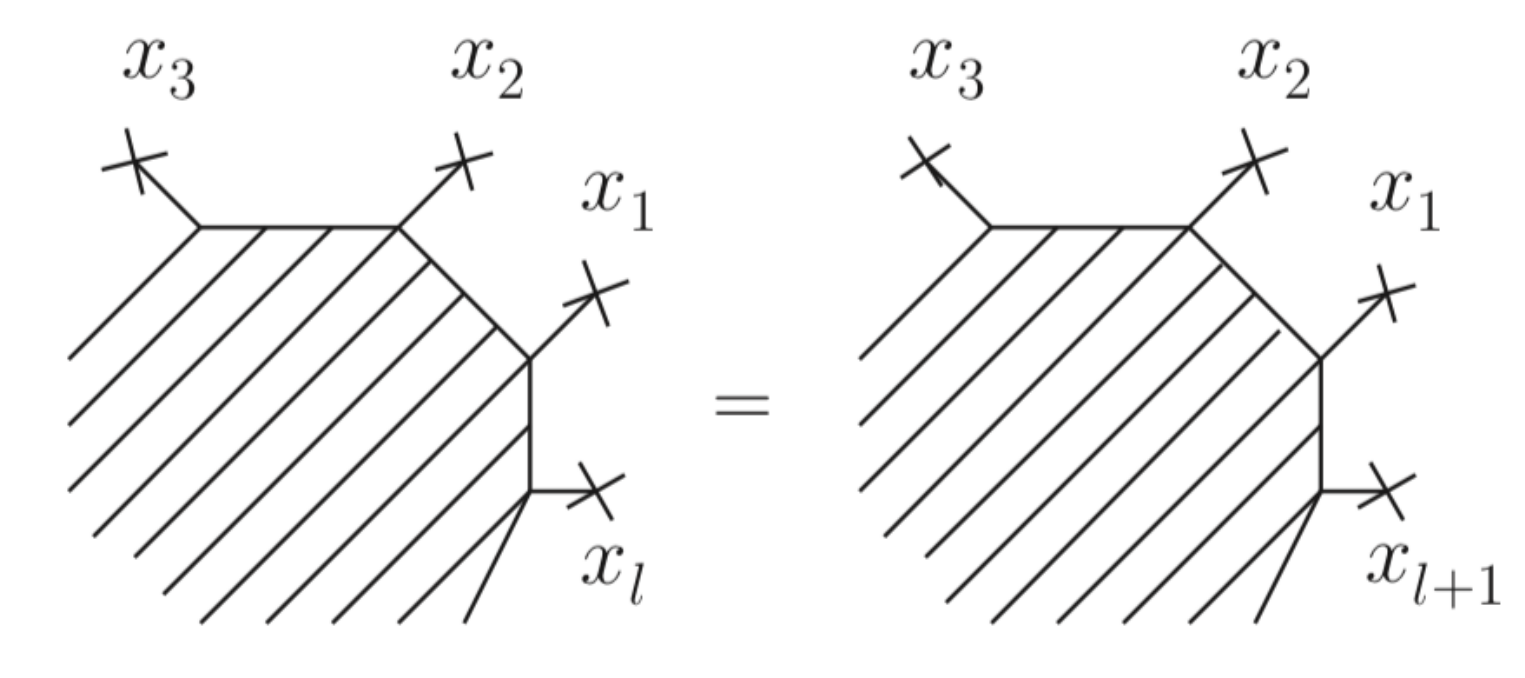}
\end{minipage}.
\label{eq:direct vertex_relation}
}

\vspace{5mm}
Finally, because the functional derivative with respect to $\rho(x)$ can be also written as 
\aln{
\frac{\delta }{\delta \rho(x)}=\int d^dy\frac{\delta(\beta U_\rho^v(y))}{\delta \rho(x)}\frac{\delta }{\delta(\beta U_\rho^{v}(y))}=-\int d^dyc_2^{}(x,y)\frac{\delta }{\delta(\beta U_\rho^{v}(y))},
}
the functional derivative of $F_2^{U_\rho^v}(x,y)$ with respect to $\rho(z)$ becomes  
\aln{
&\frac{\delta F_2^{U_\rho^v}(x,y)}{\delta\rho(z)}\bigg|_{\rho(x)=n}^{}=\int d^dw\int d^dw'F_2^{U_\rho^v}(x,w)c_3^{}(w,z,w')F_2^{U_\rho^v}(w',y),
}
where we have used Eqs. (\ref{eq:inverse relation}) and (\ref{eq:three point relation}). 
It is  graphically represented by 
\aln{
\frac{\delta }{\delta\rho(z)}
\begin{minipage}{12cm}
\includegraphics[width=12cm]{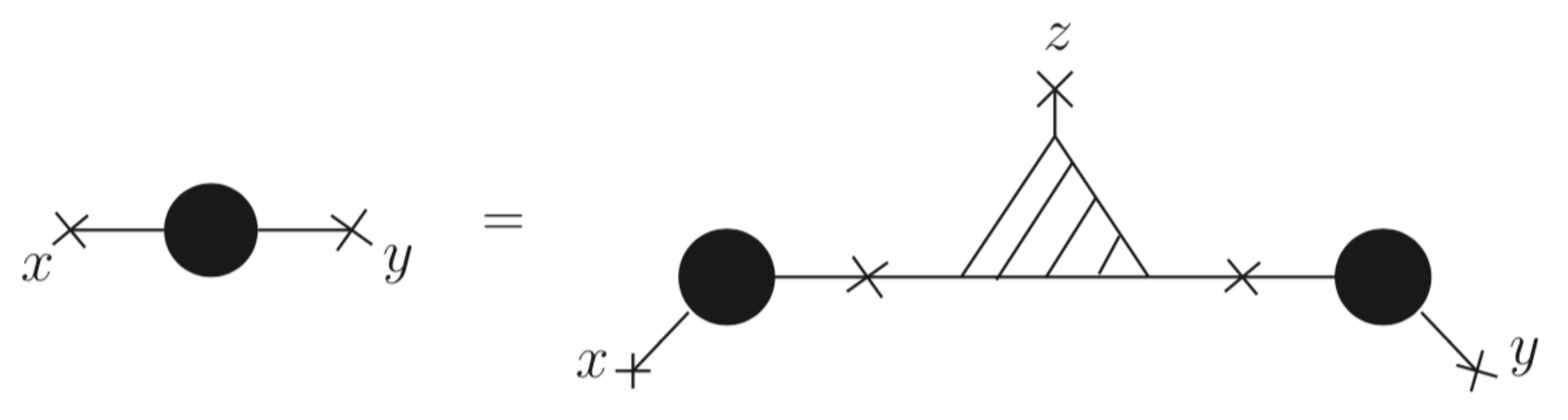}
\end{minipage}.
\label{eq:derivative of F2}
}
\\
\vspace{20mm}
Now we calculate $\Delta c_l^{}(x_1^{},\cdots,x_1^{})$ and represent them graphically.
\\
$(0)$ {\bf Zero-point $l=0$ 1PI function (Helmholtz free energy):}
\aln{\Delta (-\beta \Gamma)=-\frac{n^2}{2}\int d^dx\int d^dy\beta \delta v(x-y) -\frac{1}{2}
\begin{minipage}{3cm}
\includegraphics[width=3cm]{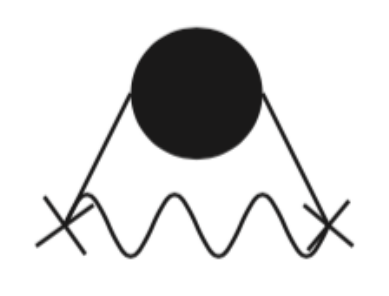}
\end{minipage}.
}
\\
$(1)$ {\bf One-point $l=1$ 1PI function (density):}
\aln{\Delta c_1^{}(x)&=-n\int d^dy\beta\delta v(x-y)
-\frac{1}{2}\begin{minipage}{6cm}
\includegraphics[width=6cm]{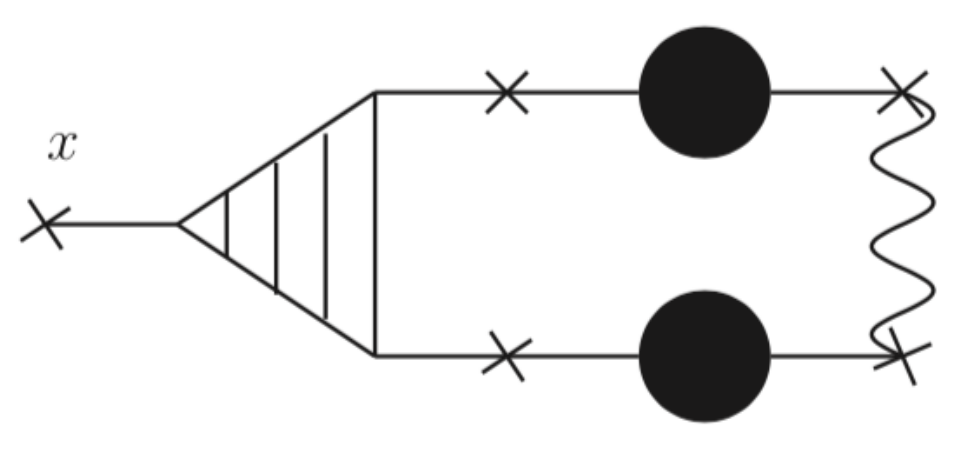}
\end{minipage}
\nn
&=-n\int d^dy\beta\delta v(x-y)-\frac{1}{2}\int d^dy \int d^dz\int d^dw\int d^dw' c_3^{}(x,w,w')F_2^{}(y,w)F_2^{}(z,w')\beta\delta v(y-z).
\label{eq:correction for direct 1}
} 
\\
$(2)$ {\bf Two-point $l=2$ direct correlation function:}
\aln{\Delta c_2^{}(x,y)= & -\beta \delta v(x-y)
-\frac{1}{2}\begin{minipage}{6cm}
\includegraphics[width=6cm]{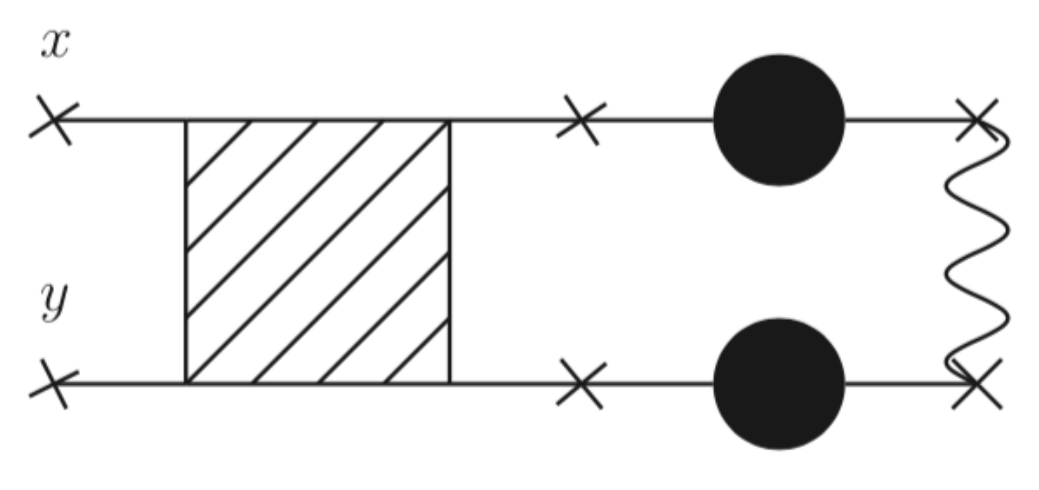}
\end{minipage}
\nn
&-\begin{minipage}{12cm}
\includegraphics[width=12cm]{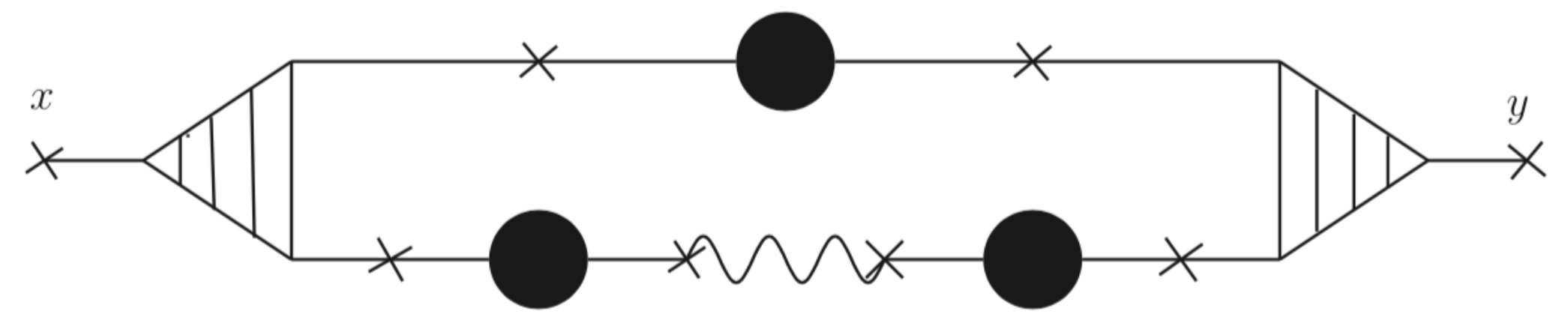}
\end{minipage}.
}
The graphical representation is read as \footnote{
One may wonder if the last two diagrams consisting of $c_4$ and $(c_3 F_2 c_3)$  
can be combined with a single term $(c_2)^4 F_4$. But the coefficients are different
and they cannot. It is because, in Eq. (\ref{eq:correction for direct 1}), 
the structure of the legs associated with the 3-point
functions  $c_3$ are asymmetric, i.e., two legs contain $F_2$ terms (the blobs) 
but the last one is amputated and there is no $F_2$ term and, when taking derivatives
with respect to $\rho$, an asymmetry between legs appears.
It is also true in the following higher order correlation functions.  
}
\aln{
&\Delta c_2^{}(x,y)= 
-\beta\delta v(x-y) \nn
& -\frac{1}{2}\int d^dz \int d^dz'\int  d^dw \int d^dw'c_4^{}(x,y,w,w')F_2^{}(z,w)F_2^{}(z',w')\beta\delta v(z-z')
\nn
&- \int
d^dz \int d^dz' \int d^dw \int d^dw'  \int d^dX  \int d^dX' 
c_3^{}(x,w,X)F_2^{}(X,z)\beta\delta v(z-z') \nn
& \hspace{50mm} \times F_2^{}(z',X') F_2^{}(w,w')c_3^{}(w',X',y) .
}
\\
$(3)$ {\bf $l$-point 1PI vertex ($l\geq 4$):}\\
In general, $\Delta c_{l}^{}(x_1^{},\cdots,x_{l}^{})$ can be systematically obtained by taking the functional derivatives of the $(l-1)$-th diagrams and using Eqs. (\ref{eq:direct vertex_relation}) and (\ref{eq:derivative of F2}).  
The $l=3$ and $l=4$ cases are given in \ref{correction:l=4 1PI}. 
\\
\subsection{DRGEs for 1PI correlation functions}
We now summarize the differential equations that the Legendre transformed correlation functions
satisfy.
By substituting the corrections $\Delta c_l$
into Eq.(\ref{eq: PDF for direct correlation}) and performing the Fourier transform, 
we obtain the hierarchical equations for the 1PI correlation functions.
In the following, we use $g_l^{}(k_1^{},\cdots,k_l^{})$ defined by 
\aln{\tilde{c}_l^{}(k_1^{},\cdots,k_l^{})
&= \int d^dx_1^{}\cdots\int d^dx_l^{}e^{-i\sum_{i=1}^lk_i^{}x_i^{}}c_l^{}(x_1^{},\cdots,x_l^{})
\nn
&\equiv (2\pi)^d\delta^{(d)}\left(\sum_{i=1}^lk_i^{}\right) g_l^{}(k_1^{},\cdots,k_l^{}),
\label{definition-gl}
}
in which the momentum conservation is factorized.
For the ideal gas (see \ref{app:ideal}), $g_l$'s  are constants Eq. (\ref{idealgas gl}) and do not have dependence on the external momenta.

\vspace{5mm}
For $l=0$, we have
\aln{
&
\frac{pV}{T}-N
=V\left(\frac{n^2}{2}\beta \tilde{v}(0)+\frac{1}{2d}\int\frac{d^dp}{(2\pi)^d}\frac{\beta\delta \tilde{v}(p)}{g_2^{}(p,-p)}\right),
}
where we have used $(\partial(-\beta \Gamma_v^{}[T,V,N])/ \partial\ln V) \big|_{T,N}^{}=pV/T$.
Noticing that $g_2(p,-p)=-1/\tilde{\kappa}(p,-p)$, it is equivalent to Eq. (\ref{eq: rge for potential}).

For $l=1$, we have
\aln{
&\frac{\partial g_1^{}(k)}{\partial\ln V}\bigg|_{T,N}^{}-1=n\beta \tilde{v}(0)-\frac{1}{2d}\int\frac{d^dp}{(2\pi)^d}\frac{g_3^{}(k,p,-p)}{g_2^{}(p,-p)^2}\beta\delta\tilde{v}(p),
\label{eq:direct rge 1}
}
which is equivalent to Eq. (\ref{eq:rge2}) 
since
\aln{g_1^{}(0)=\frac{1}{V}\int d^dx\frac{\delta (-\beta \Gamma[\rho])}{\delta \rho(x)}\bigg|_{T,V}^{}=\frac{\partial (-\beta\Gamma[n])}{\partial N}\bigg|_{T,V}^{}=-\beta \mu,
}
and we have
\aln{\frac{\partial g_1^{}(k)}{\partial\ln V}\bigg|_{T,N}^{}-1=n\frac{\partial (\beta \mu)}{\partial n}\bigg|_{T,N}^{}-1
=\kappa^{-1}-1
} 
from which we can easily check that Eq. (\ref{eq:direct rge 1}) coincides with Eq. (\ref{eq:rge2}).

For $l=2$, namely the response of the 1PI 2-point correlation
to a small change of the liquid density is given by 
\aln{
&\left(\frac{\partial}{\partial\ln V}\bigg|_{T,N}^{}-1
-\frac{1}{d}\sum_{i=1}^2k_i^\mu\frac{\partial}{\partial k_i^\mu}
\right) g_2^{}(k_1^{},k_2^{})=-\frac{\beta}{2d}(\delta\tilde{v}(k_1^{})+\delta\tilde{v}(k_2^{}))
-\frac{1}{2d}\int\frac{d^dp}{(2\pi)^d}\frac{g_4^{}(k_1^{},k_2^{},p,-p)}{g_2^{}(p,-p)^2}\beta\delta\tilde{v}(p)
\nn
&\h{2cm}+\frac{1}{d}\int\frac{d^dp}{(2\pi)^d}
\frac{g_3^{}(k_1^{},k_2^{}-p,p) g_3(k_2^{},k_1^{}+p,-p)
}{g_2^{}(p,-p)^2g_2^{}(k_1^{}+p,k_2^{}-p)}\beta\delta\tilde{v}(p). 
}
This can be rewritten as
\aln{
&\left(\frac{\partial}{\partial\ln n}\bigg|_{T,N}^{} +1
+\frac{1}{d}\sum_{i=1}^2k_i^\mu\frac{\partial}{\partial k_i^\mu}
\right) g_2^{}(k_1^{},k_2^{})=  \frac{\beta}{2d}(\delta\tilde{v}(k_1^{})+\delta\tilde{v}(k_2^{})) \nn
&
+ \frac{1}{2d}\int\frac{d^dp}{(2\pi)^d}
g_4^{}(k_1^{},k_2^{},p,-p)  \tilde{\kappa}^{}(p,-p)^2 
\beta\delta\tilde{v}(p)
\nn
& 
+ \frac{1}{d}\int\frac{d^dp}{(2\pi)^d} 
|g_3^{}(k_1^{},p,-k_1^{}-p)|^2 
\tilde\kappa^{} (p,-p)^2  \tilde\kappa^{} (k_1^{}+p,-k_1^{}-p) 
\beta\delta\tilde{v}(p). 
\label{1PI-DRGW l=2}
}
The DRGEs for $l=3$ and $l=4$ point 1PI vertices are given in \ref{app:1PIvertices}. 
Besides, in \ref{app:virial}, we also show that our DRGEs correctly reproduce the results of Mayer's cluster expansion by solving them iteratively with respect to density.    

In Eq. (\ref{1PI-DRGW l=2}), we can see an advantage of using 1PI quantities to Eq. (\ref{eq:rge3}). 
Even when higher-point vertex functions such, $g_3^{}$ and
$g_4^{}$, are replaced by the ideal gas vertices, long-range correlations in the multi-point correlation functions $F_3^{}$ or $F_4^{}$ can be at least partially taken into account through a product of two-point function $F_2^{}(x,y)$. 
Indeed, if we replace $g_3^{}$ and $g_4^{}$ by the ideal gas vertices
and set $k_2=-k_1$, Eq. (\ref{1PI-DRGW l=2}) becomes 
(we write $\kappa(k)=\kappa(k,-k)$ and $g_2^{}(k)=g_2^{}(k,-k)$)
\aln{
&\left(\frac{\partial}{\partial\ln n}\bigg|_{T,N}^{} + 1
+\frac{1}{d}k_1^\mu\frac{\partial}{\partial k_1^\mu}
\right) g_2^{}(k_1^{})  = \left(\frac{\partial} {\partial n} n 
+\frac{1}{d}k_1^\mu\frac{\partial}{\partial k_1^\mu}
\right)g_2(k_1^{})
\nn
& =  \frac{ \beta}{d} \delta\tilde{v}(k_1^{})  
- \frac{1}{dn}\int\frac{d^dp}{(2\pi)^d}
{\kappa}^{}(p)^2 
\beta\delta\tilde{v}(p)
+ \frac{1}{dn}\int\frac{d^dp}{(2\pi)^d}
{\kappa}^{} (p)^2  {\kappa}^{} (k_1^{}+p) 
\beta\delta\tilde{v}(p). 
}
It is a closed equation for the two-point correlation function. 
Validity of the approximation to replace multi-point vertices by the ideal gas needs to be
checked by studying the DRGEs for these vertices. 
Further details are studied in a separate paper. 

\section{Summary and Discussion} \label{sec:summary}

In this paper, we proposed a new formulation of statistical mechanic of classical liquid based on a scale transformation method. 
Scale transformations generate an analogue equation to the Ward-Takahashi identity of scale transformations in QFTs; and consequently we have obtained the density renormalization group equations (DRGEs). 
The set of equations describes response of various physical quantities and correlation functions to a change of the liquid density. 
The response itself depends on the density. 
Thus if we can integrate the equations from low to high density, 
 we can accumulate the effects of finite density, which corresponds to
a resummation of quantum effects in the renormalization group method in QFT. 

The DRGEs are a set of differential equations which contain multi-point correlation functions. 
It is similar to the BBGKY hierarchy. 
Namely the equations must be appropriately closed at some orders. 
Hence, our next necessary step is to introduce reasonable approximations for higher order correlation functions. 
A simple but physically reasonable approximation is to replace higher order (more than 2-point functions) 1PI vertices by those of the ideal gas (times a density-dependent function). 
It will be reasonable because in this approximation multiple-correlation effects of the liquid can be partially taken into account. 
This approximation will be systematically improved by slightly taking nonlocal effects of multiple-point 1PI vertices. 
Also it is interesting to see how we can perform resummation of the  virial expansion  by solving DRGEs. 
We will investigate these issues in separate papers. 

\section*{Acknowledgements} 
We would like to thank Yoshio Kuramoto for useful discussions. 
This work of SI is supported in part by Grants-in-Aid for Scientific
Research (No.\ 16K05329) and (No.\ 18H03708) 
from the Japan Society for the Promotion of Science.
The work of KK is supported by the Grant-in-Aid for JSPS Research Fellow, Grant Number 17J03848. 

\appendix
\def\thesection{Appendix \Alph{section}}
\section{Correlation Functions of the Ideal Gas} \label{app:ideal}
In order to confirm the consistency of the DRGEs in Eq. (\ref{eq: PDE1}), 
we calculate the correlation functions for the ideal gas.
In this case, the correlation functions must satisfy 
\aln{d\left(-\frac{\partial}{\partial(\beta \mu)}\bigg|_{T,V}^{}+\frac{\partial}{\partial\ln V}\bigg|_{\mu,V}^{}+l+\frac{1}{d}\sum_{i=1}^lx_i^\mu\partial_{i\mu}^{}
\right)F_l^{}(x_1^{},\cdots,x_l^{})
=0\label{eq:ideal SD} , 
}
which we will explicitly check in the following.

The grand canonical partition function with the external source $U(x)$ can be evaluated as
\aln{\Xi_\text{ideal}^{}[U]&=\sum_{N=0}^\infty\frac{z(\mu)^N}{N!}\int_V d^dx_1^{}\cdots \int d^dx_N^{}\exp(\beta \sum_{i=1}^N U(x_i^{}))
=\exp\left(z(\mu)\int d^dxe^{\beta U(x)}\right) 
\nn
&\therefore\ -\beta W_{\text{ideal}}^{}[U] =\log \Xi_\text{ideal}^{}[U]=z(\mu)\int d^dxe^{\beta U(x)}.
}
Thus, the correlation functions are given by 
\aln{
F_l^{}(x_1^{},\cdots,x_l^{})=n\delta^{(d)}(x_2^{}-x_1^{})\delta^{(d)}(x_3^{}-x_1^{})\cdots\delta^{(d)}(x_l^{}-x_1^{}),
}
where $n=z(\mu)=(2\pi mT)^{d/2}$. 
Thus, we have 
\aln{F_l^{}((1+\epsilon)x_1^{},(1+\epsilon)x_2^{},\cdots,(1+\epsilon)x_l^{})
&=n\delta^{(d)}((1+\epsilon)(x_2^{}-x_1^{}))\delta^{(d)}((1+\epsilon)(x_3^{}-x_1^{}))\cdots\delta^{(d)}((1+\epsilon)(x_l^{}-x_1^{}))
\nn
&=(1-(l-1)d\epsilon)F_l^{}(x_1^{},\cdots,x_l^{}),
\nn
\therefore \ \sum_{i=1}^l x_i^\mu\partial_{i\mu}^{}&
F_l^{}(x_1^{},\cdots,x_l^{})=-(l-1)dF_l^{}(x_1^{},\cdots,x_l^{}),
}
from which we can easily check that the LHS of Eq. (\ref{eq:ideal SD}) vanishes. 

Next, let us consider the Legendre transformation. 
In the ideal gas case, we can easily find the minimum of $-\beta W[U]$, and the resultant free energy is
\aln{-\beta F[\rho]&=\underset{U}{\text{Min}}(-\beta W[U]-\beta\int d^dx U(x)\rho(x))
\nn
&=\int d^dx\left[ \rho(x)- \rho(x)\log\left(\frac{\rho(x)}{(2\pi mT)^{d/2}}\right)\right],
}
from which we obtain the following direct correlation functions:
\aln{
&c_1^{}(x)=-\log\left(\frac{n}{(2\pi mT)^{d/2}}\right),\ 
c_2^{}(x,y)=-\frac{\delta^{(d)}(x-y)}{n},\ \cdots,
\nn
&c_l^{}(x_1^{},\cdots,x_l^{})=\frac{(-1)^{l+1}(l-2)!}{n^{l-1}}\delta^{(d)}(x_1^{}-x_l^{})\delta^{(d)}(x_2^{}-x_l^{})\cdots\delta^{(d)}(x_{l-1}^{}-x_l^{}).
\label{eq: correlation functions for ideal} 
}
Thus, the $\log V$ derivative in Eq. (\ref{eq: PDF for direct correlation}) gives $d(l-1)c_l^{}(x_1^{},\cdots,x_l^{})$ which is canceled by the third term in the LHS. 

For the ideal gas, $g_l^{}(k_1^{},\cdots,k_l^{})$'s defined by Eq. (\ref{definition-gl}) are given by constants
\aln{
g_l(k_1^{},\cdots,k_2^{}) = \frac{(-1)^{l+1} (l-2)!}{n^{l-1}}
\label{idealgas gl}
}
and do not have dependence on the external momenta. 

\section{DRGE for  the four-point correlation function} \label{app-F4}
The change of four-point correlation function $\Delta \tilde{F}_4$ is given by
\aln{
\Delta & \tilde{F}_4^{}(k_1^{},k_2^{},k_3^{},k_4^{})=
\begin{minipage}{14cm}
\includegraphics[width=14cm]{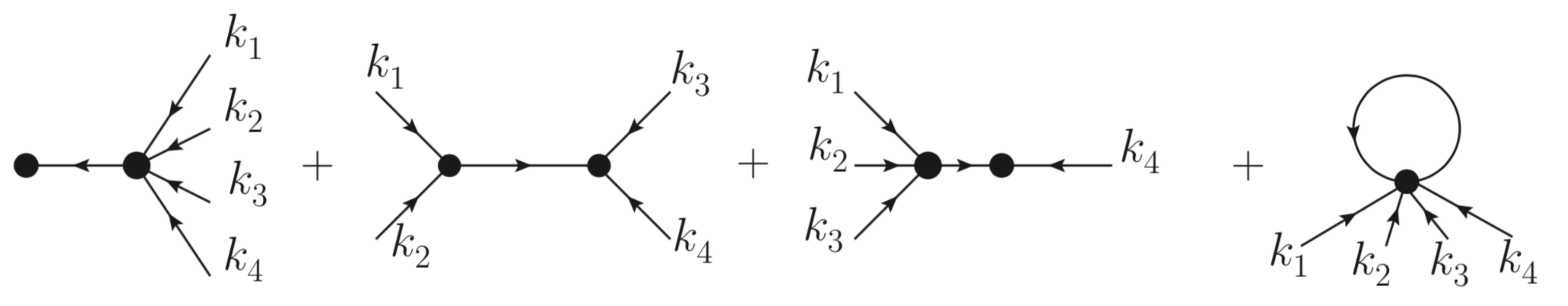}
\end{minipage}
\nn
=&-\frac{5!}{5!}\int \frac{d^dp}{(2\pi)^d}\beta\delta\tilde{v}(p)\tilde{F}_1^{}(p)\tilde{F}_5^{}(p,k_1^{},k_2^{},k_3^{},k_4^{})
\nn
&-\frac{3!3!2}{(3!)^22!}\int \frac{d^dp}{(2\pi)^d}\beta\delta\tilde{v}(p)\bigg[\tilde{F}_3^{}(k_1^{},k_2^{},-p)\tilde{F}_3^{}(p,k_3^{},k_4^{})
+\tilde{F}_3^{}(k_1^{},k_3^{},-p)\tilde{F}_3^{}(p,k_2^{},k_4^{})
\nn
&\h{4cm}+\tilde{F}_3^{}(k_1^{},k_4^{},-p)\tilde{F}_3^{}(p,k_2^{},k_3^{})
\bigg]
\nn
&-\frac{4!2!}{4!2!}\int \frac{d^dp}{(2\pi)^d}\beta\delta\tilde{v}(p)\bigg[\tilde{F}_4^{}(k_1^{},k_2^{},k_3^{},-p)\tilde{F}_2^{}(p,k_4^{})
+\tilde{F}_4^{}(k_4^{},k_1^{},k_2^{},-p)\tilde{F}_2^{}(p,k_3^{})
\nn
&\h{4cm}+\tilde{F}_4^{}(k_3^{},k_4^{},k_1^{},-p)\tilde{F}_2^{}(p,k_2^{})
+\tilde{F}_4^{}(k_2^{},k_3^{},k_4^{},-p)\tilde{F}_2^{}(p,k_1^{})
\bigg]
\nn
&-\frac{6\cdot 5\cdot 4\cdot 3}{6!}\int \frac{d^dp}{(2\pi)^d}\beta\delta\tilde{v}(p)\tilde{F}_6^{}(k_1^{},k_2^{},k_3^{},k_4^{},p,-p).
}
\\
From this, we obtain 
the DRGE for the four-point correlation function;
\aln{
&
{\cal{D}}\tilde{\lambda}_4^{}(k_1^{},k_2^{},k_3^{},k_4^{})
=-\beta\delta\tilde{v}(0)n\tilde{\lambda}_5^{}(k_1^{},k_2^{},k_3^{},k_4^{},0)
\nn
& 
- \bigg[ \tilde{\lambda}_3^{}(k_1^{},k_2^{},k_3^{}+k_4^{}) \tilde{\lambda}_3^{}(k_1^{}+k_2^{},k_3^{},k_4^{})
 \ \Re{ \left[ \beta \delta\tilde{v}(k_1^{}+k_2^{}) \right]  }
 +(k_1^{}\leftrightarrow k_3^{})+(k_1^{}\leftrightarrow k_4^{})+(k_1^{}\leftrightarrow k_3^{})
\bigg]
\nn
&-\tilde{\lambda}_4^{}(k_1^{},k_2^{},k_3^{},k_4^{})
\sum_{i=1}^4 \delta\tilde{v}(k_i^{})\tilde{\kappa}(k_i^{},-k_i^{})
-\frac{1}{2}\int\frac{d^dp}{(2\pi)^d} \beta\delta\tilde{v}(p) \tilde{\lambda}_6^{}(k_1^{},k_2^{},k_3^{},k_4^{},p,-p),
\label{eq:rge4}
}
where $\Re$ is the real part and 
$(k_a^{}\leftrightarrow k_b^{})$  
denotes  interchanging the momenta, $k_a$ and $k_b$.

\section{Towards solving DRGEs} \label{app-solve}
In this appendix, we will briefly discuss an approach to solve the DRGEs. 

There are two difficulties in solving the DRGEs.
The first one is a mixture with higher-point correlation functions,
and some approximations are necessary to close the hierarchical equations. 
Another is the momentum integration, which originates in the loop diagrams. 
The second difficulty can be avoided by noticing that various integrals have similar forms; 
integrants always contain  $\delta \tilde{v}(p)$. 
Thus, we can regard a special set of integrals as couplings that govern the system.
One of the most important examples  is given by the following integral
\aln{\kappa_I^{}\equiv \frac{1}{2d}\int\frac{d^dp}{(2\pi)^d} \beta\delta\tilde{v}(p) \kappa(p,-p) .
}
It is interpreted as a "coupling" of the liquid system as well as $\kappa(0)=\kappa_T.$ 
Other quantities appearing in the integrals of Eqs. (\ref{eq:rge2}), (\ref{eq:rge2}), and (\ref{eq:rge4}) at zero external
momenta $k_i=0$  are related to $\kappa_I$ as follows; 
\aln{
&\frac{1}{2d}\int\frac{d^dp}{(2\pi)^d} \beta\delta\tilde{v}(p) \tilde{\lambda}_3^{}(p,-p,0)
=\frac{\partial (n \kappa_I^{})}{\partial(\beta \mu)},
\\
&\frac{1}{2d}\int\frac{d^dp}{(2\pi)^d} \beta\delta\tilde{v}(p) \tilde{\lambda}_4^{}(p,-p,0,0)=\frac{\partial^2 (n\kappa_I^{})}{\partial(\beta \mu)^2},
\\
&\frac{1}{2d}\int\frac{d^dp}{(2\pi)^d} \beta\delta\tilde{v}(p) \tilde{\lambda}_5^{}(p,-p,0,0,0)=\frac{\partial^3 (n\kappa_I^{})}{\partial(\beta \mu)^3} .
}
In deriving these equations, we used the relation
\aln{
\frac{\partial }{\partial(\beta \mu)}\tilde{\lambda}_l^{}(k_1^{},\cdots ,k_l^{})\bigg|_{V,T}^{}=\tilde{\lambda}_{l+1}^{}(k_1^{},
\cdots ,k_l^{},0) ,
\label{eq:coupling relation}
}
which can be proved by using the fact that the chemical potential $\mu$ and the zero mode of $U(x)$
can be identified. In other words, the relation
\aln{
\frac{\partial}{\partial (\beta \mu)} = \int dy \frac{\partial}{\partial (\beta U(y))}  
}
is satisfied, and then Eq. (\ref{eq:coupling relation}) is derived. 
%

By using these relations in Eq. (\ref{eq: rge for potential}) and Eq. (\ref{eq:rge2}), we get the following equations:
%
\aln{&-n+\frac{p}{T}
=\frac{1}{2}\beta\tilde{v}(0)n^2-n \kappa_I^{},
\label{DRGE0}
\\
&
- \kappa_T n
+n
=\beta\tilde{v}(0)n^2 \kappa_T -\frac{\partial(n \kappa_I^{})}{\partial(\beta \mu)}\bigg|_{V,T}^{},
\label{DRGE1}
}
Setting the external momenta $k_i^{}=0$ and using these relations for $l \ge 2$, e.g., in Eq. (\ref{eq:rge3}), we get a similar equation
\aln{&
-\tilde{\lambda}_3^{}
+\tilde{\kappa}
=\beta\tilde{v}(0)n\tilde{\lambda}_3^{}
+\tilde{\kappa}^2\beta\tilde{v}(0)
-\frac{\partial^2 (n \kappa_I^{})}{\partial (\beta\mu)^2}\bigg|_{V,T}^{}.
\label{DRGE2}
}
But it is not independent from Eq.  since Eq. (\ref{DRGE2}) can be derived by taking a $(\beta \mu)$ derivative of 
Eq. (\ref{DRGE1}). 
Eq. (\ref{DRGE0}) and Eq. (\ref{DRGE1}) are not independent either, and there is a single 
independent equation for vanishing external momenta. 
Thus another relation between $\kappa_T$ and $\kappa_I$ is necessary to solve the equation. 
%

An independent equation can be obtained, e.g., by multiplying $\beta \delta \tilde{v}(k)$ on Eq. (\ref{eq:rge3}) 
with $k_1=-k_2=k$ and integrating over $k$. 
Then, defining two new couplings by the following integrals
\aln{
\kappa_{I2}^{}  \equiv \frac{n}{2d^2}\int\frac{d^dp}{(2\pi)^d} (\beta\delta\tilde{v}(p) \kappa(p,-p))^2, 
}
and 
\aln{
\lambda_{4}^{} \equiv \frac{1}{(2d)^2}\int \frac{d^dp}{(2\pi)^d} \int \frac{d^dk}{(2\pi)^d}
\beta\delta\tilde{v}(p) \beta\delta\tilde{v}(k) \lambda_4^{}(p,-p,k,-k) ,
}
we get the following equation
\aln{&
\left( - \frac{\partial}{\partial (\beta \mu)} +1 \right) (n \kappa_I) = 
\beta \tilde{v}(0) n \frac{\partial(n \kappa_I)}{\partial (\beta \mu)} 
 - n\kappa_{I2}^{} - n\lambda_4^{} .
}
In this way we can generate independent differential equations, but at the same time 
more new couplings are  introduced and we need some approximations to
close the equations; it is the destiny of the hierarchical equations and 
further investigations are left for future publications.

\section{Corrections to $l=3, 4$ 1PI vertices} \label{correction:l=4 1PI}
The correction to the $l=3$ 1PI vertex $\Delta c_3^{}(x,y,z)$ is graphically given by
\aln{\Delta c_3^{}(x,y,z)=&-\frac{1}{2}
\begin{minipage}{6cm}
\includegraphics[width=6cm]{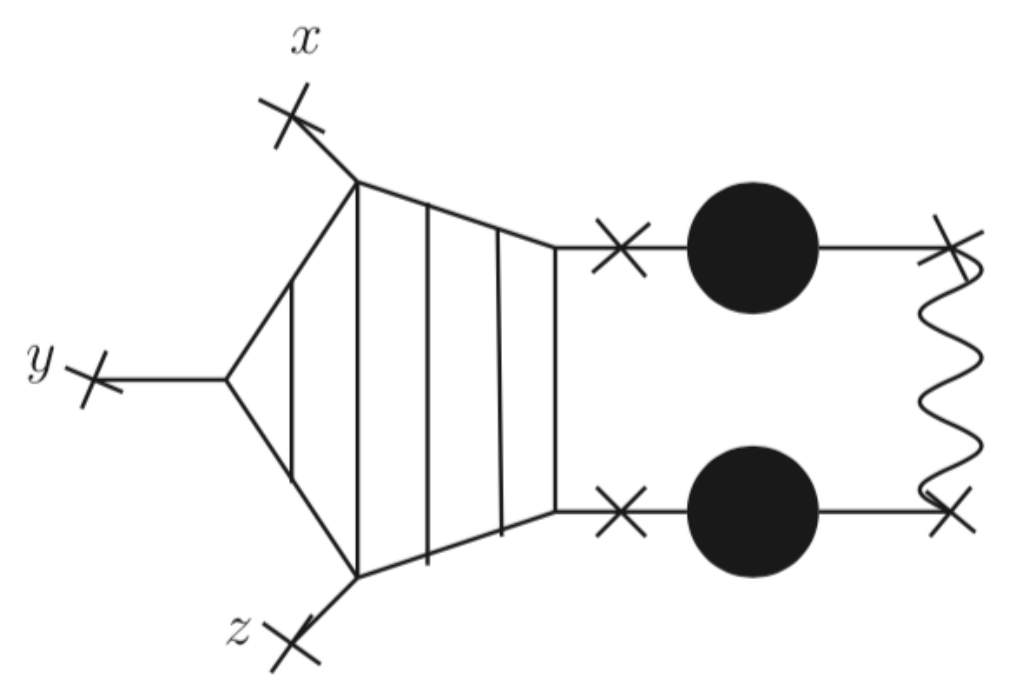}
\end{minipage}
\nonumber
\\
&-\frac{1}{2}\left[\begin{minipage}{10cm}
\includegraphics[width=10cm]{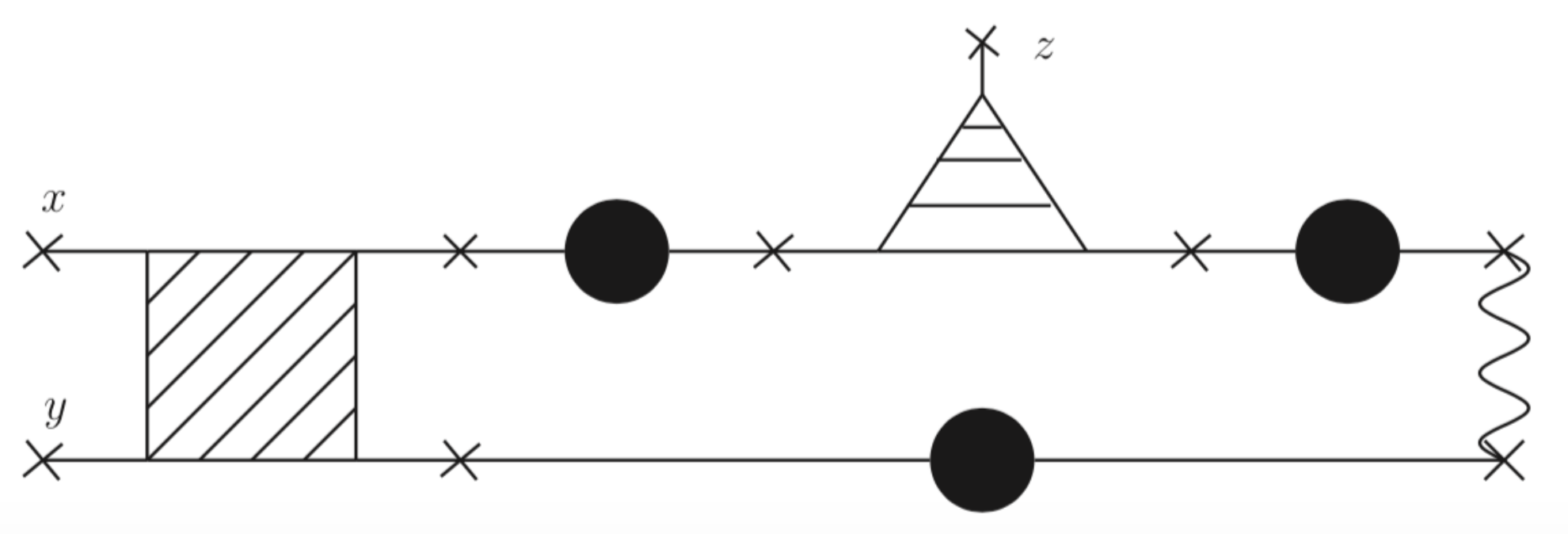}
\end{minipage}
\right]_s^{}
\nonumber
}
\aln{\hspace{3.5cm}&-\frac{1}{2}\left[\begin{minipage}{12cm}
\includegraphics[width=12cm]{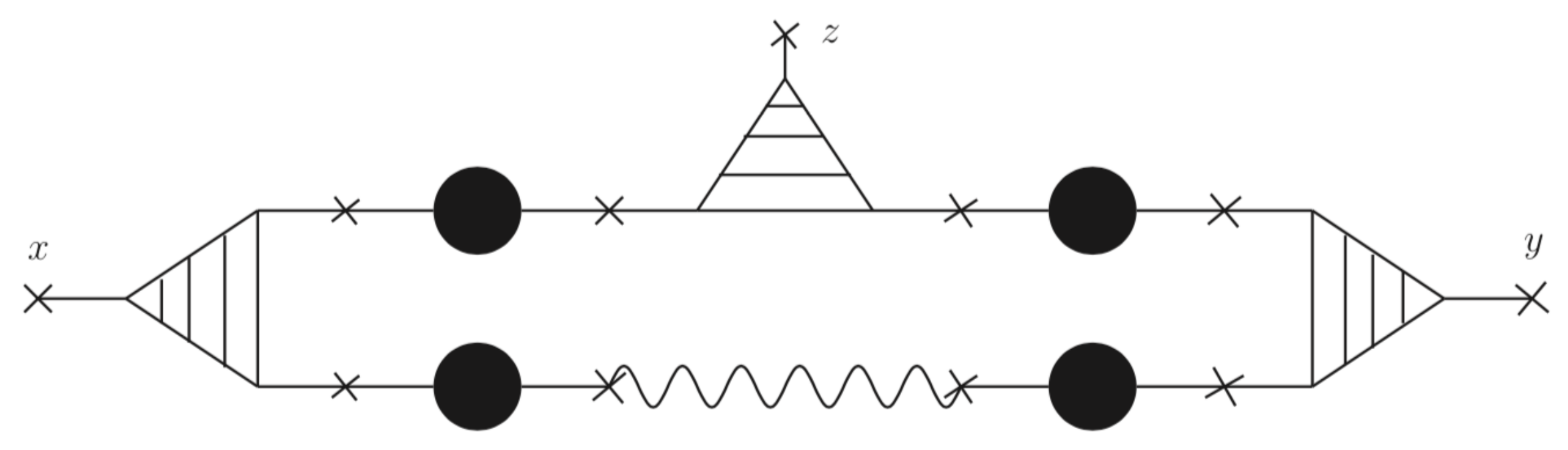}
\end{minipage}
\right]_s^{},
\label{eq:c3_diagrams}
}
where $[ f(x,y,z) ]_s^{}$ is a symmetrization over 3 variables;
\aln{
[ f(x_1, x_2, x_3) ]_s^{} = \sum_{\sigma \in S_3^{}} f(x_{\sigma(1)}^{}, x_{\sigma(2)}^{}, x_{\sigma(3)}^{}).
}
The graphical representation is read as
\aln{
{\cal{D}} c_3^{}(x,y,z)=&-\frac{1}{2}\left(\prod_{i=1}^2\int d^ds_i^{}\right)c_5^{}(x,y,z,s_1^{},s_2^{})
 (F_2*\delta v *F_2)(s_1, s_2)
\nn
-\frac{1}{2} 
\bigg[\left(\prod_{i=1}^4 \int d^ds_i^{}\right)&c_4^{}(x,y,s_1^{},s_4^{})F_2^{}(s_1^{},s_2^{})c_3^{}(s_2^{},z,s_3^{})
 (F_2*\delta v *F_2)(s_3, s_4)
\bigg]_s
\nn
- \frac{1}{2}
\bigg[\left(\prod_{i=1}^6 \int d^ds_i^{}\right)&c_3^{}(x,s_1^{},s_6^{})F_2^{}(s_1^{},s_2^{})c_3^{}(s_2^{},z,s_3^{})
F_2^{}(s_3^{},s_4^{})c_3^{}(s_4^{},y,s_5^{}) 
 (F_2*\delta v *F_2)(s_5, s_6)
\bigg]_s^{}.
\label{eq: DRG for c3}
}
The correction to the $l=4$ 1PI vertex $\Delta c_4^{}(x,y,z,w)$ is graphically given by
\aln{\Delta c_4^{}(x,y,z,w)=&-\frac{1}{2}
\begin{minipage}{6cm}
\includegraphics[width=6cm]{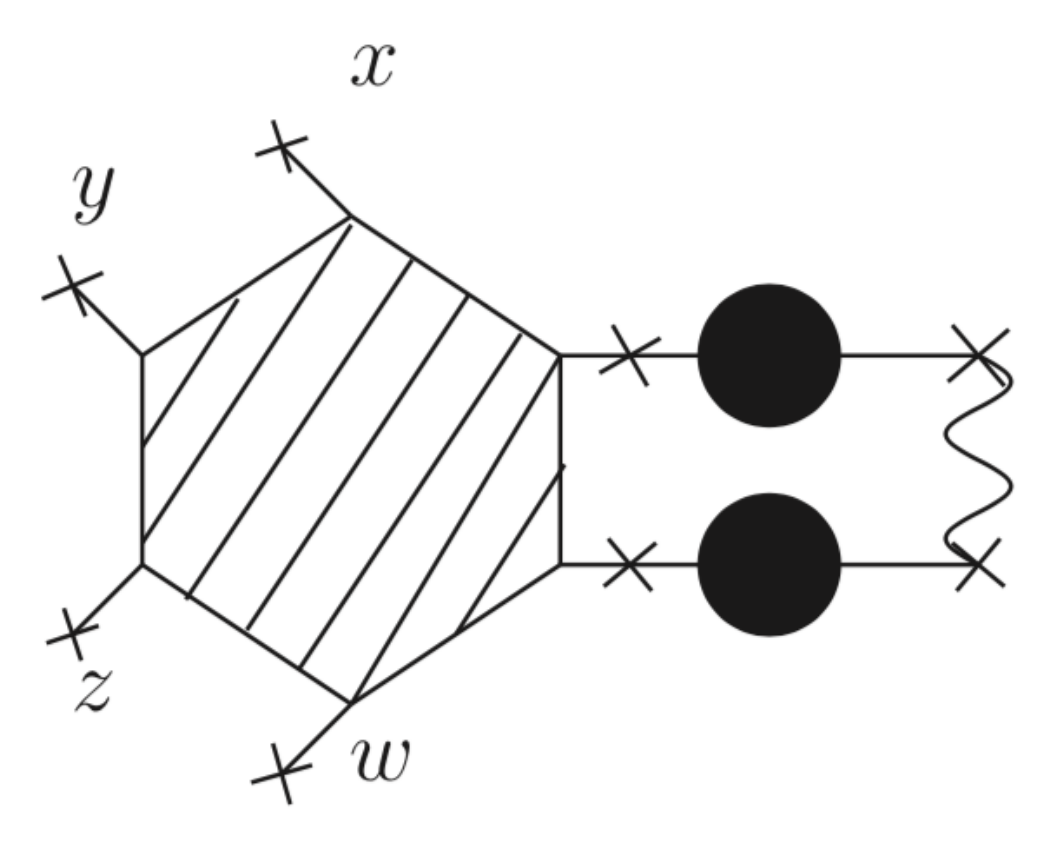}
\end{minipage}
\nonumber
}
\aln{
\h{3.5cm}&-\frac{1}{3!}\left[\begin{minipage}{8cm}
\includegraphics[width=8cm]{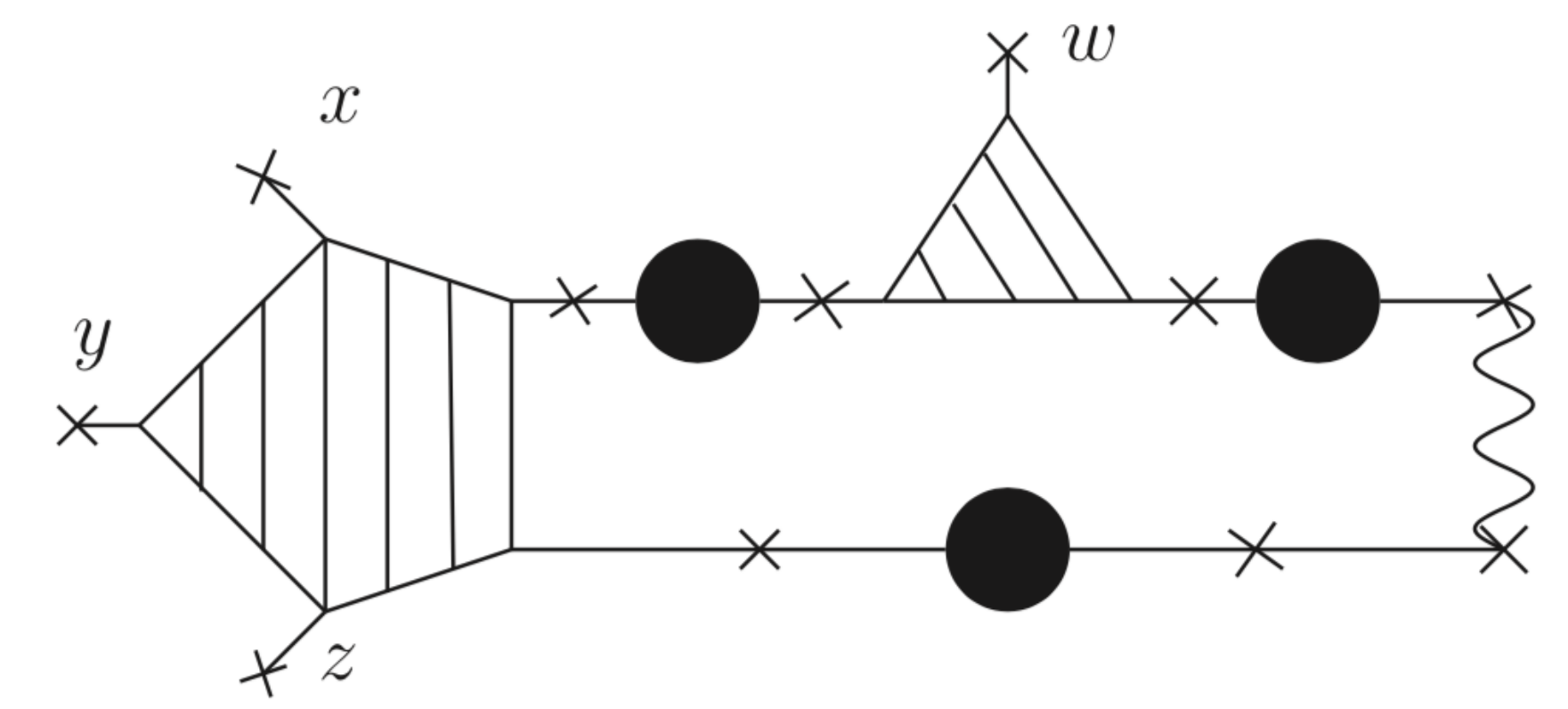}
\end{minipage}
\right]_s^{}
\nn
&-\frac{1}{2!2!2!}\left[\begin{minipage}{10cm}
\includegraphics[width=10cm]{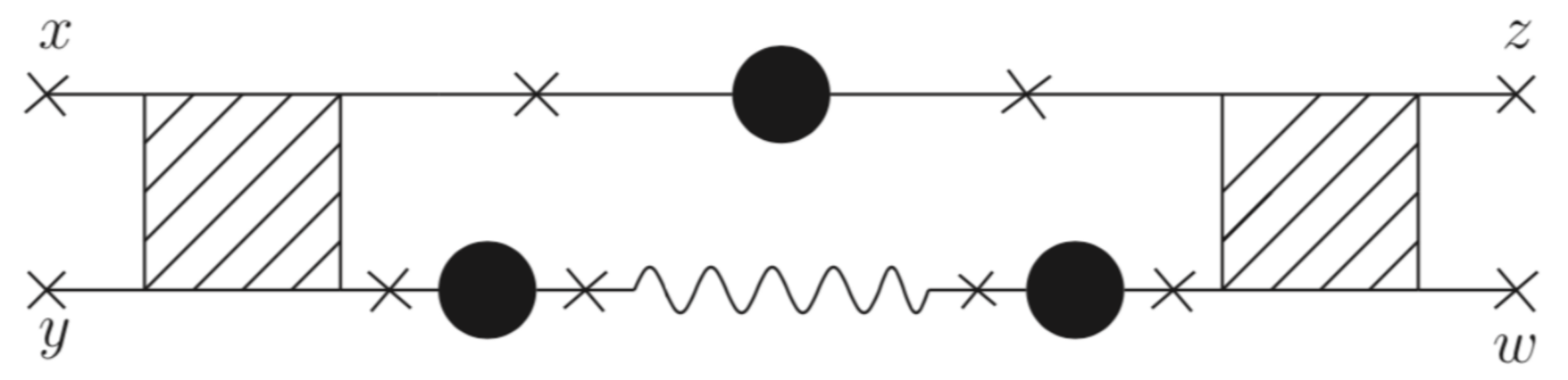}
\end{minipage}
\right]_s^{}
\nn
&-\frac{1}{2}\left[\begin{minipage}{10cm}
\includegraphics[width=10cm]{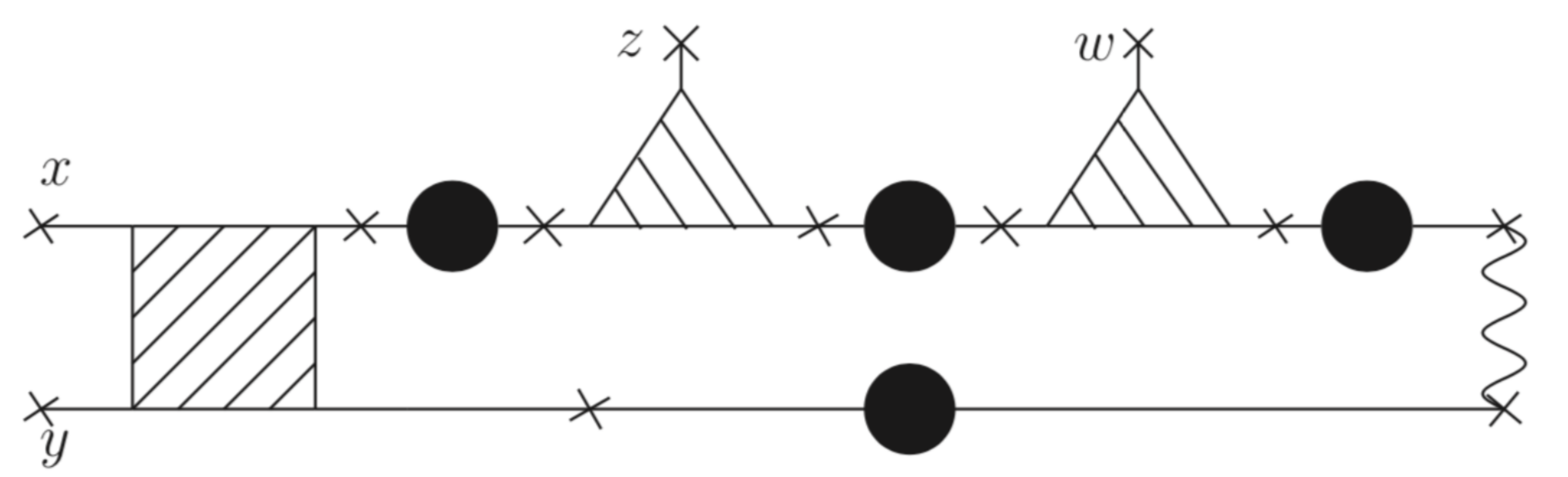}
\end{minipage}
\right]_s^{}
\nn
&-\frac{1}{2!2!}\left[\begin{minipage}{8cm}
\includegraphics[width=8cm]{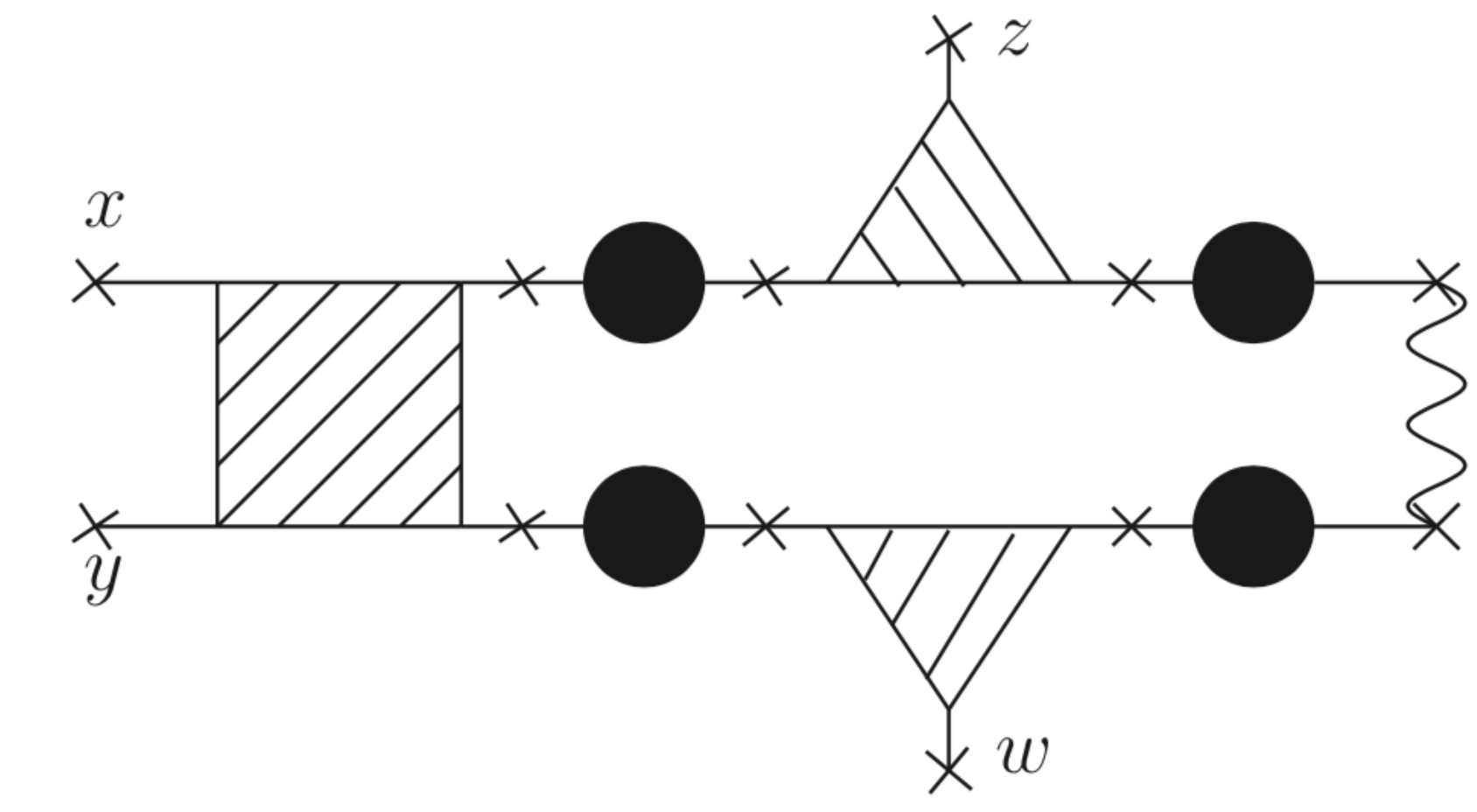}
\end{minipage}
\right]_s^{}
\nn
&-\frac{1}{2}\left[\begin{minipage}{12cm}
\includegraphics[width=12cm]{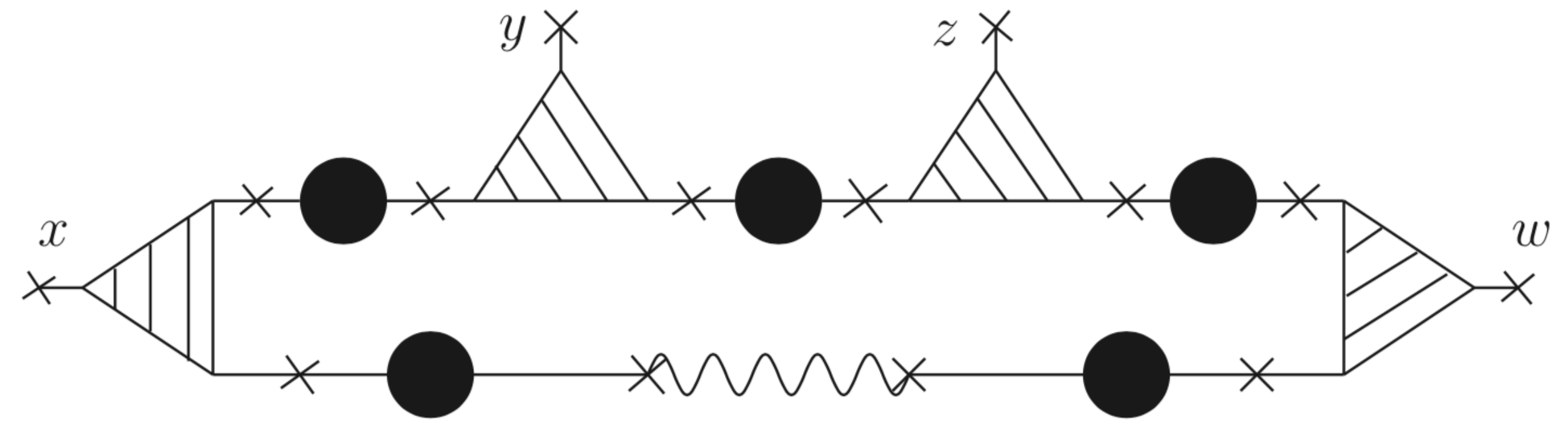}
\end{minipage}
\right]_s^{},
\label{eq:c4_diagrams}
}
where $[f(x_1^{},x_2^{},x_3^{},x_4^{})]_s^{}$ means a symmetrization over 4 variables;
\aln{[f(x_1^{},x_2^{},x_3^{},x_4^{})]_s^{}\equiv \sum_{\sigma\in S_4^{}}f(x_{\sigma(1)}^{},x_{\sigma(2)}^{},x_{\sigma(3)}^{},x_{\sigma(4)}^{}).
}
The graphical representation is read as 
\aln{
&{\cal{D}} c_4^{}(x,y,z,w)=-\frac{1}{2}\left(\prod_{i=1}^2\int d^ds_i^{}\right)c_6^{}(x,y,z,w,s_1^{},s_2^{})
 (F_2*\delta v *F_2)(s_1, s_2)
\nn
&-\frac{1}{6} \bigg[\left(\prod_{i=1}^4 \int d^ds_i^{}\right)c_5^{}(x,y,z,s_1^{},s_4^{})F_2^{}(s_1^{},s_2^{})c_3^{}(s_2^{},z,s_3^{})
 (F_2*\delta v *F_2)(s_3, s_4)
\bigg]_s
\nn
&-\frac{1}{8}  \bigg[\left(\prod_{i=1}^4 \int d^ds_i^{}\right)c_4^{}(x,y,s_1^{},s_4^{})F_2^{}(s_1^{},s_2^{})c_4^{}(s_2^{},z,w,s_3^{})
 (F_2*\delta v *F_2)(s_3, s_4)
\bigg]_s
\nn
&- \frac{1}{2}  \bigg[\left(\prod_{i=1}^6 \int d^ds_i^{}\right)
c_4^{}(x,y,s_1^{},s_6^{})F_2^{}(s_1^{},s_2^{})c_3^{}(s_2^{},z,s_3^{})F_2^{}(s_3^{},s_4^{})c_3^{}(s_4^{},w,s_5^{})
 (F_2*\delta v *F_2)(s_5, s_6) \bigg]_s
\nn
&- \frac{1}{4}  \bigg[\left(\prod_{i=1}^6 \int d^ds_i^{}\right)
c_4^{}(x,y,s_1^{},s_6^{})F_2^{}(s_1^{},s_2^{})c_3^{}(s_2^{},z,s_3^{})
 (F_2*\delta v *F_2)(s_3, s_4)
c_3^{}(s_4^{},z,s_5^{})F_2^{}(s_5^{},s_6^{}) \bigg]_s
\nn
&-\frac{1}{2}\bigg[\left(\prod_{i=1}^{8}\int d^ds_i^{}\right)
c_3^{}(x,s_1^{},s_{8}^{})F_2^{}(s_1^{},s_2^{})c_3^{}(s_2^{},y,s_3^{})F_2^{}(s_3^{},s_4^{})c_3^{}(s_4^{},z,s_5^{})F_2^{}(s_5^{},s_6^{})c_3^{}(s_6^{},w,s_7^{}) \nn
& \hspace{30mm} 
 (F_2*\delta v *F_2)(s_7, s_{8})
\bigg]_s .
\label{eq: DRG for c4}
}     


\section{DRGEs for $l=3, 4$ 1PI vertices} \label{app:1PIvertices}
The DRGE for $l=3$ 1PI vertex can be read from Eq. (\ref{eq:c3_diagrams}) and given by
\aln{
&\left( \frac{\partial}{\partial\ln n}\bigg|_{T,N}^{} +2
+\frac{1}{d}\sum_{i=1}^3k_i^\mu\frac{\partial}{\partial k_i^\mu}
\right) g_3^{}(k_1^{},k_2^{},k_3^{}) =\left( \frac{1}{n} \frac{\partial}{\partial n} n^2+\frac{1}{d}\sum_{i=1}^3k_i^\mu\frac{\partial}{\partial k_i^\mu}\right)g_3^{}(k_1^{}, k_2^{}, k_3^{})
\nn
&= \frac{1}{2d}\int\frac{d^dp}{(2\pi)^d}\frac{g_5^{}(k_1^{},k_2^{},k_3^{},p,-p)}{g_2^{}(p,-p)^2}\beta\delta\tilde{v}(p)
\nn
& -\frac{1}{d}\bigg[\int\frac{d^dp}{(2\pi)^d}\frac{g_4^{}(k_1^{},k_2^{},p,k_3^{}-p)g_3^{}(-k_3^{}+p,k_3^{},-p)}{g_2^{}(-k_3^{}+p,k_3^{}-p)g_2^{}(p,-p)^2}\beta\delta\tilde{v}(p)
+(k_1^{}\leftrightarrow k_3^{})+(k_2^{}\leftrightarrow k_3^{})
\bigg]
\nn
&+ \frac{1}{d}\bigg[\int\frac{d^dp}{(2\pi)^d}\frac{g_3^{}(k_1^{},p,-k_1^{}-p)g_3^{}(k_2^{},k_1^{}+p,k_3^{}-p)g_3^{}(-k_3^{}+p,-p,k_3^{})}{g_2^{}(k_1^{}+p,-k_1^{}-p)g_2^{}(-k_3^{}+p,k_3^{}-p)g_2^{}(p,-p)^2}\beta\delta\tilde{v}(p)+(k_1^{}\leftrightarrow k_2^{})+(k_2^{}\leftrightarrow k_3^{})
\bigg].
}
For $l=4$ 1PI vertex, the DRGE is given by
\aln{
&\left( \frac{\partial}{\partial\ln n}\bigg|_{T,N}^{} + 3
+\frac{1}{d}\sum_{i=1}^4k_i^\mu\frac{\partial}{\partial k_i^\mu}
\right) g_4^{}(k_1^{},k_2^{},k_3^{},k_4^{}) =\left( \frac{1}{n^2} \frac{\partial}{\partial n} n^3 +\frac{1}{d}\sum_{i=1}^4k_i^\mu\frac{\partial}{\partial k_i^\mu}\right)g_4^{}(k_1^{}, k_2^{}, k_3^{},k_4^{})
\nn
&= \frac{1}{2d}\int\frac{d^dp}{(2\pi)^d}\frac{g_6^{}(k_1^{},k_2^{},k_3^{},k_4^{},p,-p)}{g_2^{}(p,-p)^2}\beta\delta\tilde{v}(p)\nn
& - \frac{1}{d}\bigg[\int\frac{d^dp}{(2\pi)^d}\frac{g_5^{}(k_1^{},k_2^{},k_3^{},p,k_4^{}-p)g_3^{}(-k_4^{}+p,k_4^{},-p)}{g_2^{}(-k_4^{}+p,k_4^{}-p)g_2^{}(p,-p)^2}\beta\delta\tilde{v}(p)
+(k_1^{}\leftrightarrow k_4^{})+(k_2^{}\leftrightarrow k_4^{})+(k_3^{}\leftrightarrow k_4^{})\bigg]
\nn
&+\frac{1}{d}\bigg[\int\frac{d^dp}{(2\pi)^d}\frac{g_4^{}(k_1^{},k_2^{},p,-k_1^{}-k_2^{}-p)g_3^{}(k_1^{}+k_2^{}+p,k_3^{},k_4^{}-p)g_3^{}(-k_4^{}+p,k_4^{},p)}{g_2^{}(k_1^{}+k_2^{}+p,-k_1^{}-k_2^{}-p)g_2^{}(-k_4^{}+p,k_4^{}-p)g_2^{}(p,-p)^2}\beta\delta\tilde{v}(p)
\nn
&\quad+(k_1^{}\leftrightarrow k_3^{})+(k_1^{}\leftrightarrow k_4^{})+(k_2^{}\leftrightarrow k_3^{})+(k_2^{}\leftrightarrow k_4^{})+(k_1^{}\leftrightarrow k_3^{}\ \&\ k_2^{}\leftrightarrow k_4^{})
\nn
&\quad +\int\frac{d^dp}{(2\pi)^d}\frac{g_4^{}(k_1^{},k_2^{},p,-k_1^{}-k_2^{}-p)g_3^{}(k_1^{}+k_2^{}+p,k_4^{},k_3^{}-p)g_3^{}(-k_3^{}+p,k_3^{},p)}{g_2^{}(k_1^{}+k_2^{}+p,-k_1^{}-k_2^{}-p)g_2^{}(k_3^{}-p,-k_3^{}+p)g_2^{}(p,-p)^2}\beta\delta\tilde{v}(p)
\nn
&\quad+(k_1^{}\leftrightarrow k_3^{})+(k_1^{}\leftrightarrow k_4^{})+(k_2^{}\leftrightarrow k_3^{})+(k_2^{}\leftrightarrow k_4^{})+(k_1^{}\leftrightarrow k_3^{}\ \&\ k_2^{}\leftrightarrow k_4^{})
\nn
&\quad+\int\frac{d^dp}{(2\pi)^d}\frac{g_4^{}(k_1^{},k_2^{},p,-k_1^{}-k_2^{}-p)g_3^{}(k_1^{}+k_2^{}+p,k_3^{},k_4^{}-p)}{g_2^{}(k_1^{}+k_2^{}+p,-k_1^{}-k_2^{}-p)g_2^{}(-k_4^{}+p,k_4^{}-p)^2g_2^{}(p,-p)}
 g_3^{}(-k_4^{}+p,k_4^{},-p)\beta\delta\tilde{v}(-k_4^{}+p)
\nn
&\quad+(k_1^{}\leftrightarrow k_3^{})+(k_1^{}\leftrightarrow k_4^{})+(k_2^{}\leftrightarrow k_3^{})+(k_2^{}\leftrightarrow k_4^{})+(k_1^{}\leftrightarrow k_3^{}\ \&\ k_2^{}\leftrightarrow k_4^{})\bigg]
\nn
&- \frac{2}{d}\bigg[\int\frac{d^dp}{(2\pi)^d}\frac{g_3^{}(k_1^{},p,-k_1^{}-p)g_3^{}(k_1^{}+p,k_2^{},-k_1^{}-k_2^{}-p)}
{g_2^{}(k_1^{}+p,-k_1^{}-p)g_2^{}(k_1^{}+k_2^{}+p,-k_1^{}-k_2^{}-p)g_2^{}(-k_4^{}+p,k_4^{}-p)} \nn
& \hspace{10mm} \times \frac{g_3^{}(k_1^{}+k_2^{}+p,k_3^{},k_4^{}-p)g_3^{}(-k_4^{}+p,k_4^{},-p)}{g_2^{}(p,-p)^2} 
\beta\delta\tilde{v}(p)
\nn
&\quad+(k_1^{}\leftrightarrow k_2^{})+(k_1^{}\leftrightarrow k_3^{})+(k_2^{}\leftrightarrow k_4^{})+(k_3^{}\leftrightarrow k_4^{})+(k_1^{}\leftrightarrow k_2^{}\ \&\ k_3^{}\leftrightarrow k_4^{})\bigg]. 
\label{eq:direct rge 4}
}
Here, $(k_1^{}\leftrightarrow k_2^{})$ represents interchanging $k_1$ and $k_2$.
%
The sign
 $(k_1^{}\leftrightarrow k_2^{}\ \&\ k_3^{}\leftrightarrow k_4^{})$  means 
interchanging both of $(k_1^{}, k_1^{})$ and $(k_3^{}, k_4^{})$.
%
%

\section{Virial Expansion} \label{app:virial}
Here, we show that Mayer's cluster expansions are systematically reproduced by solving the DRGEs iteratively.  
Before the detailed calculation, let us explain a relation between the direct correlation functions and the virial expansions. 
The virial expansion of the equation of state (EOS) for liquid/vapor is defined by 
\aln{\frac{p}{T}=n+B_2^{}(T)n^2+B_3^{}(T)n^3+\cdots,
}
where $\{B_i^{}(T)\}$ are virial coefficients which are traditionally calculated based on Mayer's cluster expansion.  
For simple liquid, they are explicitly given by   
\aln{
& B_2^{}(T)=-\frac{1}{2}\int d^dx_1^{} f_{12}^{},\label{eq: second virial coefficient} 
\\
& B_3^{}(T)=-\frac{1}{3}\int d^dx_1^{}\int d^dx_2^{}f_{12}^{}f_{23}^{}f_{31}^{},\ \cdots,
\label{eq: third virial coefficient} 
}
where 
\aln{f_{12}^{}\equiv e^{-\beta v(x_1^{}-x_2^{})}-1
}
is Mayer's $f$ function. 
From the point of view of DRGEs, these coefficients are related to two-point direct correlation function $c_2^{}(x)$ as follows.   
By integrating the inverse relation Eq. (\ref{eq:inverse relation}),we obtain 
\aln{\tilde{c}_2^{}(0)\equiv \int d^dxc_2^{}(x)=-\frac{1}{\tilde{F}_2^{}(0)}=-\frac{1}{nT}\frac{\partial p}{\partial n}\bigg|_{T}^{},
} 
where we have used the thermodynamical relation Eq. (\ref{eq: compressibility by F2}) for the isothermal compressibility. 
Thus, if we expand $c_2^{}(x)$ with respect to  the density as 
\aln{c_2^{}(x-y)=-\frac{\delta^{(d)}(x-y)}{n}+
\sum_{m=1}^{\infty} n^{m-1} c_2^{(m)}(x-y),
\label{expand-c2}
}
we obtain the following relations between the virial coefficients $B_{m+1}$ and $c_2^{(m)}$: 
\aln{
B_{m+1}^{}(T)=-\frac{1}{m+1}\int d^dx \ c_2^{(m)}(x).
\label{eq: relation between c2 and virial}
}
Two-point correlation function $F_2^{}(x)$ is  expanded as
\aln{
&F_2^{}(x-y)=n\delta^{(d)}(x-y)+ \sum_{m=1}^{\infty} n^{m+1}  F_2^{(m)}(x-y) .
\label{eq: density expansion of F2}
}
Since $F_2^{(i)}$ is an inverse of $c_2^{(j)}$  as in Eq. (\ref{eq:inverse relation}),
they are related to each other:
\aln{ 
& F_2^{(1)}(x)=c_2^{(1)}(x), \nn
& F_2^{(2)}(x)=c_2^{(2)}(x)+(c_2^{(1)}*c_2^{(1)})(x), \nn
& F_2^{(3)}(x)=c_2^{(3)}(x)+ 2 (c_2^{(1)}*c_2^{(2)})(x) 
+ (c_2^{(1)}*c_2^{(1)}* c_2^{(1)})(x). 
\label{eq: F2 and c2}
}
Higher $F_2^{(p)}$ $p \ge 4$ can be similarly written in terms of $c_2^{(q)}$ with $q \le p$.

\vspace{10mm}

The 1PI vertices $c_l^{}(x_1, \cdots, x_l)$ for higher $l \ge 3$ 
can be  similarly expanded as 
\aln{
&c_l^{}(x_1^{},\cdots,x_l^{})=\frac{(-1)^{l+1}(l-1)!}{n^{l-1}}\prod_{i=2}^l\delta^{(d)}(x_1^{}-x_i^{})
+ \sum_{m=1}^{\infty} n^{m+1-l}  c_l^{(m)}(x_1^{},\cdots,x_l^{})  .
\label{eq: density expansion of cl}
}
Note that  for  the $c_l^{(m)}$ terms $m \le l-2$  are singular at $n=0$. 
This indicates that these coefficients should vanish:
\aln{
c_l^{(1)} = c_l^{(2)} = \cdots = c_l^{(l-2)} = 0. 
\label{lower-c-vanish}
}
In the following, we explicitly check the properties  for $l=3$ and $l=4$.
Here we explain why it should be so. 
From Eq. (\ref{eq: density derivative}), the following relation 
\aln{
\int d^d y  \ c_{l+1}(x_1, \cdots, x_l, y) =  \frac{\partial c_{l}(x_1, \cdots, x_l)}{\partial n}
\label{cl-cl-relation}
}
is satisfied. 
The 1PI vertices for the ideal gas in Eq. (\ref{eq: correlation functions for ideal}) actually satisfy the relation. 
This ideal gas contribution is the only non-analytic term in $c_2(x)$ in Eq. (\ref{expand-c2}). 
Since other terms of $c_2(x)$ obtained by the Virial expansion is an analytic function of $n$, the relation of Eq. (\ref{cl-cl-relation}) shows that an integral of higher order correlation functions is also analytic with respect to $n$. For example, we have the relations:
\aln{
& \int d^dy \ c_3^{(1)}(x_1, x_2, y) = 0,  \\
& \int d^dy \ c_3^{(2)}(x_1, x_2, y) = c_2^{(2)}(x_1, x_2), \\
& \int d^dy \ c_3^{(3)}(x_1, x_2, y) = 2 c_2^{(3)}(x_1, x_2).
}
This does not completely guarantee that $c_3^{(1)} = 0$, but it is required by analyticity of the virial expansion. 
Namely, if we impose that the virial expansion of the two-point correlation function in an $x$-dependent background $\rho(x)$ is analytic with respect to $\rho(x)$, $c_3^{(1)}(x_1, x_2, x_3)$ must vanish since it becomes singular at $\rho(x)=0$. 
Similarly, $c_4^{(1)} = c_4^{(2)} =0$ is expected, and so is the relation $c_l^{(1)} = c_l^{(2)} = \cdots = c_l^{(l-2)} = 0$. 
\vskip1cm
Let us now  solve the DRGEs iteratively with respect to  the density $n$.  
By substituting the expansions Eq. (\ref{eq: density expansion of cl}) into the DRGEs and picking up the leading order contributions, we obtain  the following relations: for $l=2$ we have
\aln{&\hat{s}c_2^{(1)}(x_1^{}-x_2^{})
=-(1+ c_2^{(1)}(x_1^{}-x_2^{}))\beta \delta v(x_1^{}-x_2^{})
\nn
&-\frac{1}{2}\int d^ds\int d^ds' c_4^{(1)}(x_1^{}-x_2^{},0,s,s')\beta\delta v(s-s')
-2\int d^ds  c_3^{(1)}(x_1^{}-x_2^{},0,s)\beta \delta v(s) .
\label{eq:c_2_leading}
}
For $l=3$,
\aln{
\left(d+\hat{s}\right) c_3^{(1)}(x_1^{},x_2^{},x_3^{})=&- c_3^{(1)}(x_1^{},x_2^{},x_3^{})\beta\left[\delta v(x_1^{}-x_2^{})+\delta v(x_2^{}-x_3^{})+\delta v(x_3^{}-x_1^{})
\right]
\nn
&-\int d^ds c_4^{(1)}(x_1^{},x_2^{},x_3^{},s)\beta\left[\delta v(x_1^{}-s)+\delta v(x_2^{}-s)+\delta v(x_3^{}-s)
\right].
}
For $l=4$,
\aln{
&\left(2d+\hat{s}\right) c_4^{(1)}(x_1^{},x_2^{},x_3^{},x_4^{})
\nn
=&-c_4^{(1)}(x_1^{},x_2^{},x_3^{},x_4^{})\beta [\delta v(x_1^{}-x_2^{})+\delta v(x_1^{}-x_3^{})+\delta v(x_1^{}-x_4^{})+\delta v(x_2^{}-x_3^{})+\delta v(x_2^{}-x_4^{})+
\delta v(x_3^{}-x_4^{})]
\nn
&-\frac{1}{2}\int d^ds_1^{}\int d^ds_2^{}c_6^{(1)}(x_1^{},x_2^{},x_3^{},x_4^{},s_1^{},s_2^{})\beta \delta v(s_1^{}-s_2^{})
\nn
&-\int d^dsc_5^{(1)}(x_1^{},x_2^{},x_3^{},x_4^{},s)\beta [\delta v(s-x)+\delta v(s-y)+\delta v(s-z)+\delta v(s-w)] .
\label{eq:c4 leading}
}
An important property of these equations is that the RHS of the $l$-th equation does not depend on lower-point correlation functions $c_j^{(1)}\ (j\leq l-1)$ (at least up to $l=4$). 
In particular, there is no $c_2^{(1)}$ dependence for the $l\geq 3$ equations.  
As a result, these equations (except for $l=2$)  have trivial solutions, $c_l^{(1)}=0\ (l\geq 3)$,
which is consistent with Eq. (\ref{lower-c-vanish}) and the analyticity of virial expansions. 
Putting $c_3^{(1)}=0,\ c_4^{(1)}=0$, Eq. (\ref{eq:c_2_leading}) becomes simplified as
\aln{
\hat{s} c_2^{(1)}(x_1-x_2) = -(1+c_2^{(1)}(x_1-x_2)) \beta v(x_1-x_2),
}
and we have the following solution:  
\aln{
& c_2^{(1)}(x_1^{}-x_2^{})=e^{-\beta v(x_1^{}-x_2^{})}-1=f_{12}^{},  
\label{eq:c_2_leading_2}
}
where we used a boundary condition such that $c_2^{(1)}$ should vanish for $v(x)$=0. 
From Eq. (\ref{eq: relation between c2 and virial}), we can see that this solution correctly reproduces the second virial coefficient Eq. (\ref{eq: second virial coefficient}).  

\vskip1cm

Next, by concentrating on the next order terms, we have 
\aln{
&\left(-d+\hat{s}\right)c_2^{(2)}(x_1^{}-x_2^{})=- F_2^{(2)}(x_1^{}-x_2^{})\beta \delta v(x_1^{}-x_2^{})-2c_2^{(1)}(x_1^{}-x_2^{})(\beta\delta v* c_2^{(1)})(x-y)
\nn
&-\frac{1}{2}\int d^ds\int d^ds' c_4^{(2)}(x_1^{}-x_2^{},0,s,s')\beta\delta v(s-s')
-2\int d^ds c_3^{(2)}(x-y,0,s)\beta \delta v(s)
\nn
=&-c_2^{(2)}(x-y)\beta \delta v(x-y)-\beta \delta v(x-y)( c_2^{(1)}* c_2^{(1)})(x-y)
-2 c_2^{(1)}(x-y)(\beta\delta v* c_2^{(1)})(x-y)
\nn
&-\frac{1}{2}\int d^ds\int d^ds'c_4^{(2)}(x_1^{}-x_2^{},0,s,s')\beta\delta v(s-s')
-2\int d^ds c_3^{(2)}(x_1^{}-x_2^{},0,s)\beta \delta v(s),
\label{eq: c2 next order}
} 
for $l=2$, 
\aln{\hat{s} c_3^{(2)}(x_1^{},x_2^{},x_3^{})=&-[ c_3^{(2)}(x_1^{},x_2^{},x_3^{})+ c_2^{(1)}(x_1^{}-x_2^{}) c_2^{(1)}(x_2^{}-x_3^{})]\beta \delta v(x_3^{}-x_1^{})
\nn
&-[ c_3^{(2)}(x_1^{},x_2^{},x_3^{})+ c_2^{(1)}(x_2^{}-x_3^{})c_2^{(1)}(x_3^{}-x_1^{})]\beta \delta v(x_1^{}-x_2^{})
\nn
&-[ c_3^{(2)}(x_1^{},x_2^{},x_3^{})+ c_2^{(1)}(x_3^{}-x_1^{}) c_2^{(1)}(x_1^{}-x_2^{})]\beta \delta v(x_2^{}-x_3^{})
\nn
&-\int d^ds c_4^{(1)}(x_1^{},x_2^{},x_3^{},s)\beta  [\delta  v(x_1^{}-s)+\delta v(x_2^{}-s)+\delta v(x_3^{}-s)],
\label{eq:c3 next-leading}
}
for $l=3$, and     
\aln{&\left(d+\hat{s}\right)c_4^{(2)}(x_1^{},x_2^{},x_3^{},x_4^{})
\nn
=&-c_4^{(2)}(x_1^{},x_2^{},x_3^{},x_4^{})\beta [\delta v(x_1^{}-x_2^{})+\delta v(x_1^{}-x_3^{})+\delta v(x_1^{}-x_4^{})+\delta v(x_2^{}-x_3^{})+\delta v(x_2^{}-x_4^{})+
\delta v(x_3^{}-x_4^{})]
\nn
&-\frac{1}{2}\int d^ds_1^{}\int d^ds_2^{}c_6^{(2)}(x_1^{},x_2^{},x_3^{},x_4^{},s_1^{},s_2^{})\beta \delta v(s_1^{}-s_2^{})
\nn
&-\int d^dsc_5^{(2)}(x_1^{},x_2^{},x_3^{},x_4^{},s)\beta [\delta v(s-x)+\delta v(s-y)+\delta v(s-z)+\delta v(s-w)]
,\label{eq:c4 next-leading}
}
for $l=4$. 
Note that the $l=4$ case is essentially the same as the leading order one Eq. (\ref{eq:c4 leading}) because contributions that contain $c_2^{(1)}\times c_2^{(1)}$ also vanish. 
Thus, as well as the leading order case, this equation also allows trivial solution $c_4^{(2)}=c_5^{(2)}=c_6^{(2)}=0$, and this leads to a  following solution for $c_3^{(2)}$: 
\aln{c_3^{(2)}(x_1^{},x_2^{},x_3^{})&=(e^{-\beta v(x_1^{}-x_2^{})}-1)(e^{-\beta v(x_2^{}-x_3^{})}-1)(e^{-\beta v(x_3^{}-x_1^{})}-1)
\nn
&=f_{12}^{}f_{23}^{}f_{31}^{},
}
which corresponds to the integrand of the third virial coefficient $B_3^{}(T)$. 
%
%
The second Virial coefficient for the two-point function $c_2^{(2)}$ can be
obtained by using the relation of Eq. (\ref{cl-cl-relation}) and we have
\aln{ c_2^{(2)}(x)=c_2^{(1)}(x) (c_2^{(1)}*c_2^{(1)})(x) .
} 
It satisfies Eq. (\ref{eq: c2 next order}) with $c_4^{(2)}=0$ and $c_3^{(2)}=f_{12}^{}f_{23}^{}f_{31}^{}$. 
%

%


\end{document}